\documentclass[english]{article}
\usepackage[T1]{fontenc}
\usepackage[latin9]{inputenc}
\usepackage{geometry}
\usepackage{amsmath}
\usepackage{amssymb}
\usepackage{graphicx}

\makeatletter
\newcommand{\lyxaddress}[1]{
\par {\raggedright #1
\vspace{1.4em}
\noindent\par}
}

\makeatother

\usepackage{babel}
\begin{document}

\title{Non-adiabatic dynamics of electrons and atoms \\under non-equilibrium
conditions}

\author{L. Kantorovich}
\maketitle

\lyxaddress{Physics Department, King's College London, The Strand, London, WC2R
2LS, United Kingdom}
\begin{abstract}
An approach to non-adiabatic dynamics of atoms in molecular and condensed
matter systems under general non-equilibrium conditions is proposed.
In this method interaction between nuclei and electrons is considered
explicitly up to the second order in atomic displacements defined
with respect to the mean atomic trajectory, enabling one to consider
movement of atoms beyond their simple vibrations. Both electrons and
nuclei are treated fully quantum-mechanically using a combination
of path integrals applied to nuclei and non-equilibrium Green's functions
(NEGF) to elections. Our method is partition-less: initially, the
entire system is coupled and assumed to be at thermal equilibrium.
Then, the exact application of the Hubbard-Stratanovich transformation
in mixed real and imaginary times enables us to obtain, without doing
any additional approximations, an exact expression for the reduced
density matrix for nuclei and hence an effective quantum Liouville
equation for them, both containing Gaussian noises. It is shown that
the time evolution of the expectation values for atomic positions
is described by an infinite hierarchy of stochastic differential equations
for atomic positions and momenta and their various fluctuations. The
actual dynamics is obtained by sampling all stochastic trajectories.
It is expected that applications of the method may include photo-induced
chemical reactions (e.g. dissociation), electromigration, atomic manipulation
in scanning tunneling microscopy, to name just a few.
\end{abstract}

\section{Introduction}

There are a very large number of phenomena in physics, chemistry and
biology in which it is necessary to consider non-adiabatic dynamics
of system atoms. Prominent examples include, e.g. in biology and chemistry:
photosynthesis \cite{Brixter-Nature-2005,Tapavicza-PCCP-2011}, vision
\cite{Hayashi-Biophys-2009,Polli-Nature-2010}, photoisomerization
of rhodopsin and isorhodopsin \cite{Chung-JPCB-2012}, molecular photochemistry
of biomolecules \cite{Sobolewski-PCCP-2002}, proton \cite{Fang-JCP-1997,Varella-JCP-2006,Marx-CPC-2006}
and electron transfer \cite{Marcus-BioPhys-1985,Cotton-JCP-2014,Hammes-Schiffer-ChemRev-2010,Duncan-AnnRevPC-2007}
reactions, also between distant redox centres \cite{Stih-Science-2008}.
Non-adiabatic dynamics is often essential in energy production (photovoltaics)
\cite{Rozzi-NatComm-2013,Jailaubekov-NatMat-2013}, in photo-induced
dissociation \cite{Xu-JACS-2014,Kling-Science-2006} and isomerisation
\cite{Levine-AnnRevPC-2007} dynamics, in femtosecond chemistry \cite{Zewail-JPCA-2000,Richter-JPCL-2012},
oxygen production in comets \cite{Yao-NatComm-2017}, acceleration
of urethane and polyurethane formation due to vibrational excitation
\cite{Stensitzki-NatChem-2018}, etc. In physics non-adiabatic effects
are also widespread and may be highly important, e.g., in vibrationally
promoted electron emission from a metal surface \cite{Nahler-Science-2008},
dynamics of nanoparticles under strong laser pulses \cite{Bonafe-Nanoscale-2017},
coupling of plasmons and vibrations in nanoparticles \cite{Ahmed-ACSNano-2017},
electromigration \cite{Todorov-PRL-2001,DiVentra-PRL-2002,Yang-PRB-2003,Verdozzi-PRL-2006,Dundas-NatNano-2009,Esen-APL-2005,Sordan-APL-2005,Hoffman-APL-2008,Umeno-APL-2009,Araidai-PRB-2009,Kizuka-APE-2009,Taychatanapat-NanoLett-2007}
that can adversely affect the nanodevices due to atomic rearrangement
leading to their subsequent degradation \cite{Bode-PRL-2011,Tierney-NatNano-2011},
local heating in a conductor (e.g., in atomic wires) \cite{Smit-Nanotech-2004,Tsutsui-APL-2007,Jintao-NanoLett-2010,Jintao-PRL-2015,Brandbyge-PRB-2012,Horsfield-RPP-2006},
photoelectron spectroscopy \cite{Avigo-JPCM-2013}, radiation damage
\cite{Duffy-NIMOR-2009}, and atomic manipulation is scanning tunneling
microscopy \cite{Bartels-PRL-1997,Keeling-PRL-2005,Bellec-PRL-2010,Labidi-PRB-2012},
to name just a few. 

A considerable number of theoretical tools have been developed over
the years to tackle this kind of problems where dynamics of both electrons
and nuclei is considered simultaneously. These methods can be crudely
divided into two big classes: (i) wave-functions based methods applicable
at zero temperatures, and (ii) density matrix based methods which
can be applied at any temperature. 

In the simplest mixed quantum-classical Ehrenfest approach, within
the first class of methods, the nuclei are treated classically (they
satisfy classical equations of motion) while electronic wavefunction
is evolved in time via the time-dependent Sch\"odinger equation \cite{Saita-JCP-2012}.
If transitions between different potential energy surfaces (PES),
e.g. due to an optical excitation (and after it upon relaxation),
are required to consider, the simplest strategy is offered by the
fewest switches surface hopping method \cite{Tully-JCP-1971,Tully-JCP-1990},
in which regions near conical intersections of the PESs along the
adiabatic trajectory are branched in a certain way. The advantages
and (many) disadvantages of this approximate method are critically
discussed in the reviews \cite{Tully-JCP-2012,Subotnik-AnnRevPC-2016}. 

We note that there is also a method in which the evolution of the
electronic subsystem is replaced by the dynamics of a system of fictitious
harmonic oscillators; this enables one to run molecular dynamics simulations
of non-adiabatic processes entirely classically. A \textquotedblleft quantization\textquotedblright{}
of the electronic states is added approximately. Some successes of
this method are reviewed in \cite{Miller-FaradDiss-2016}. 

An expression for the atomic forces \emph{due to electrons} is required
to couple classical equations of motion for atoms and time evolution
of the electronic wavefunction. Usually, the force on the classical
atomic degree of freedom $A$ in the quantum-classical approaches
is calculated via an expression $F_{A}^{e}=-\left\langle \psi_{t}\right|\partial_{x_{A}}H_{e}\left|\psi_{t}\right\rangle $
with $H_{e}$ being the electronic Hamiltonian (that includes interaction
with nuclei) and $\psi_{t}$ the corresponding many-electron wavefunction
\cite{Cunningham-PRB-2014,Todorov-PRB-2010,Dundas-NatNano-2009,Cunningham-BJN-2015,Jane-JCTC-2016,Tully-JCP-2012,Subotnik-AnnRevPC-2016}
(or the density matrix for the electrons \cite{vOppen-Belst-2012}).
Since only the potential energy of interaction between nuclei and
electrons, $V_{ne}$, in $H_{e}$ actually depends on the atomic positions
$x_{A}$, one gets $\partial_{x_{A}}H_{e}\equiv\partial_{x_{A}}V_{ne}$,
and so the above expression for the force then formally coincides
\cite{Zhang-PRB-2-11} with the Hellman-Feynman force normally used
in density functional calculations \cite{DiVentra-book,Kantorovich-SS-book}.
Note that a proper definition of this force is essential, especially
under non-equilibrium conditions, and not in all cases it can be assigned
simply to the gradient of the potential energy. Indeed, for instance,
at the current flow conditions (e.g., in molecular junctions and nanodevices),
when open boundary conditions are used, the number of electrons is
not well defined, and hence the potential energy. This problem is
formally solved by appealing to the original (and formally exact)
Ehrenfest equations \cite{DiVentra-PRB-2000,DiVentra-book,Todorov-JPCM-2001},
which, if written for zero temperature, are: 
\[
m_{A}\partial_{t}\left\langle x_{A}\right\rangle _{t}=\left\langle p_{A}\right\rangle _{t}\:,\quad\partial_{t}\left\langle p_{A}\right\rangle _{t}=F_{A}^{e}+F_{A}^{i}=-\left\langle \partial_{x_{A}}V_{ne}\right\rangle _{t}+F_{A}^{i}
\]
where $F_{A}^{i}$ is the contribution to the total force due to direct
interaction between atoms. In these equations the averages $\left\langle \ldots\right\rangle _{t}=\left\langle \Psi_{t}\right|\ldots\left|\Psi_{t}\right\rangle $
are assumed with the wavefunction $\Psi_{t}$ for the combined electron-nuclear
system, so replacing in the expression of the force $\Psi_{t}$ with
$\psi_{t}$ may seem, although intuitively appealing, still an approximation.
In fact, it can easily be shown that this result is exact if it is
assumed, within the model of classical nuclei, that the electron-nuclear
interaction is described by the integral $\int n_{e}(\mathbf{r})\widehat{v}_{ne}(\mathbf{r})d\mathbf{r}$,
where $n_{e}(\mathbf{r})$ is the electron density and $\widehat{v}_{ne}(\mathbf{r})$
the one-electron potential provided by the nuclei \cite{Todorov-JPCM-2001}.
Indeed, in this case this term contributes, in the Lagrangian equations
of motion (with atoms treated classically), a contribution 
\[
\int n_{e}(\mathbf{r})\left(-\partial_{x_{A}}\widehat{v}_{ne}(\mathbf{r})\right)d\mathbf{r}\equiv-\left\langle \psi_{t}\right|\partial_{x_{A}}V_{ne}\left|\psi_{t}\right\rangle 
\]
which is exactly the contribution employed in quantum-classical approaches.
Note that the atomic force thus defined has non-zero curl and hence
is not conservative, as may be anticipated \cite{Dundas-NatNano-2009,Todorov-PRB-2010}. 

The next, more sophisticated class of methods, still based on the
wavefunction treatment of the electronic subsystem, uses Gaussian
wave packets (GWP) to represent the nuclei wavefunction \cite{Worth-AdvChemPh-2002,Worth-MolPh-2008}.
The PESs in these methods are calculated ``on the fly'' which is
efficient. There are several variants of this method: trajectory surface
hopping (TSH) \cite{Tully-JCP-1971,Tully-JCP-1990,Hack-JPCA-2000},
coupled coherent states (CCS) \cite{Shalashilin-CP-2004}, \emph{ab
initio} multiple spawning (AIMS) \cite{Ben-Nun-AdvChemPh-2002}, multiconfigurational
Ehrenfest (MCE) \cite{Saita-JCP-2012}, \emph{ab initio }multiple
cloning \cite{Makhov-JCP-2014}, and variational multiconfigurational
Gaussian wavepacket (vMCG) \cite{Worth-CPL-2003,Worth-FaradDiss-2004},
the latter being more flexible than the others (the parameters of
the wavepackets are determined ``on the fly'' as well), can describe
tunneling, but it is also more expensive and numerically more difficult
to handle \cite{Worth-MolPh-2008}. The multi-configuration time-dependent
Hartree (MCTDH) \cite{Meyer-CPL-1990,Beck-PhRep-2000,Worth-IntRevPhCh-2008}
method can be considered as a generalization of the previously mentioned
methods that use Gaussian basis, as in MCTDH the nuclear basis is
more general. Although these methods, especially their generalised
variants vMCG and MCTDH, may provide (in the limit of the complete
basis set) an exact solution of the electron-nuclear time-dependent
Sch\"odinger equation, the methods are quite expensive computationally
and can only be applied to small systems (a small number of nuclear
degrees of freedom). 

E. K. U. Gross \emph{et al}. have developed a reformulation of the
exact time-dependent Sch\"odinger equation in which the wavefunction
of the combined system is factorised in the Born-Oppenheimer (BO)
form as a product of two variational functions: one for electrons,
which depends on the nuclear positions, and one for nuclei \cite{Abedi-PRL-2010,Abedi-JCP-2012,Min-PRL-2015}.
The two equations for each of the wavefunctions are coupled by a scalar
and vector potentials that are subject to some gauge conditions. The
two equations are strictly equivalent to the original Sch\"odinger
equation, and hence are not easier to solve. One advantage of this
method is based on the fact that the wavefunction is not expanded
into BO electronic wavefunctions for each electronic state, and hence
PES for each such a state does not appear. Instead,  an effective
PES is introduced (the mentioned scalar potential), which corresponds
to an\emph{ effective propagation} of the system in time. This proved
to be useful in analyzing results of the dynamics simulations. The
other advantage of this method is that it allows introducing approximations
in a more controlled way. Various approximate incarnations of this
method have been applied to a number of applications (see, e.g., \cite{Agostini-JCP-2015,Min-JPCL-2017}),
demonstrating that the method is very promising. 

Concerning density matrix based methods, a number of approaches exist
varying in underlying approximations and the cost of the calculations.
In quantum-classical Liouville equation (QCLE) method \cite{Kapral-JCP-1999,Kelly-JCP-2013,MacKerman-JPCB-2008,Kapral-JPCM-2015}
the most important degrees of freedom are treated quantum mechanically
(called ``the open system''), while the rest of the variables (``the
bath'') are treated approximately as semi-classical. The latter is
done by, first, transforming the Liouville equation using the Wigner
transform with respect to the bath variables and then making an expansion
in the power series with respect to $\hbar$. In the first order an
intuitively expected result is obtained for the transformed Liouville
operator that becomes a simple sum of (symmetrised) classical and
quantum Poisson brackets \cite{Aleksandrov-ZN-1981,Prezhdo-PRA-1997,Prezhdo-JCP-2000,Prezhdo-TCA-2006,Prezhdo-JCP-2006,Yu-LK-PRB-2007}.
This approach enables one to obtain an approximate equation of motion
for the reduced (with respect to the bath degrees of freedom) open
system density matrix; the classical variables are evolved in time
classically. In the generalized quantum master equation (GQME) method
\cite{Kelly-JCP-2013} the classical bath degrees of freedom are projected
out from the Liouville equation using Nakajima-Zwanzig projection
operators, and then the partitioned approach is applied to obtain
a self-contained equation for the reduced density matrix of the system.
This equation has the form of the first order differential equation
with an integral memory term. Then approximations are applied to the
calculation of the kernel in the memory term. The partitioned approach
assumes the density matrix at the initial time is factorised into
a direct product of independent density matrices of the system and
bath, i.e. the whole system is initially decoupled. Moreover, the
bath is assumed to be at thermal equilibrium. A more general approach
that can treat the initial system-bath coupling was developed in \cite{Zhang-JCP-2006}.
It is argued in \cite{Kelly-JCP-2013,Pfalzgraff-JPCL-2015} that if
in QCLE only short time evolutions are accessible, the main advantage
of the GQME approach is that one can access relatively longer time
scales in the dynamics. 

Another way to consider both nuclei and electrons quantum-mechanically
is based on path integrals. The most popular are two approaches, the
ring-polymer molecular dynamics (RPMD) \cite{Craig-JCP-2004,Habershon-AnnRevPhChem-2013}
and centroid molecular dynamics (CMD) \cite{Cao-JCP-1994,Jang-JCP-1999}.
In both methods the starting point is the imaginary time path integral
representation of the partition function for the nuclear system associated,
initially, with the (single) ground state PES that could be calculated,
for instance, with advanced \emph{ab initio }electronic structure
methods like density functional theory (DFT). Then the mapping between
the Hamiltonian in the Euclidean action of the path integral and that
of a ring polymer is exploited that enables one to run ``classical''
molecular dynamics simulations in extended phase space. The method
is very efficient and can be applied to systems containing hundreds
of atoms. If initially these methods were only applied to an adiabatic
dynamics on a single PES, its extensions to non-adiabatic dynamics
have also been proposed, both for RPMD \cite{Ananth-JCP-2013,Richardson-JCP-2013,Richardson-CP-2017,Kretchmer-FaradDiss-2016,Shushkov-JCP-2012}
and CMD \cite{Schwieters-JCP-1999,Liao-JPCB-2002} incarnations. The
main limitation of the above methods is that they are designed only
for equilibrium; one cannot use these methods for investigating time-dependent
and non-equilibrium phenomena, although non-equilibrium situations
have also started to be addressed \cite{Welsch-JCP-2016}. 

The methods reviewed so far were derived at different levels of theory
and using various approximations. Amongst the wavefunction based methods,
approaches based on GWPs are still computationally expensive and can
only be applied to relatively small systems. The computationally cheap
Ehrenfest based methods with hoppings have a number of shortcomings
which cannot be controlled. The method developed in the Gross' group,
although theoretically elegant, if applied directly without any approximation,
is computationally expensive; only its approximate variants can be
used to study realistic systems. A definite advantage of this method
is that it is not based on the adiabatic PES. Non-zero temperature
methods based on solving the Liouville equation are all approximate,
treating nuclei semi-classically. Finally, the path integral based
approaches cannot be directly applied to non-equilibrium phenomena.
Neither of the previously considered techniques is universal; for
instance, it is not obvious how they can be used for problems that
require open boundary conditions (e.g. to study current carrying molecular
junctions). 

A systematic approach that can be applied to a wide class of problems,
including the ones with open boundary conditions, has also been developed
called the correlated electron-ion dynamics (CEID) \cite{McEniry-EPJB-2010}.
In the initial formulation of the method \cite{Horsfield-JPCM-2004,Horsfield-JPCM-2005}
the Hamiltonian is expanded in a Taylor series with respect to the
atomic displacements $u_{A}=x_{A}-\left\langle x_{A}\right\rangle _{t}$
from the mean atomic trajectories $\left\langle x_{A}\right\rangle _{t}=\text{Tr}\left(\widehat{\rho}(t)x_{A}\right)$
(where $\widehat{\rho}(t)$ is the density matrix of the entire system
at time $t$), then various correlation functions appear corresponding
to fluctuations of positions, $u_{A}$, and momenta $\Delta p_{A}=p_{A}-\left\langle p_{A}\right\rangle _{t}$,
for which equations of motion are derived as well. This procedure
leads to an infinite hierarchy of first order differential equations
which is terminated at a certain order. In \cite{Stella_JCP-2007,Stella-JCP-2011}
an entirely new formulation of the method has been developed based
on Wigner transform, which enables one to derive the CEID equations
up to an arbitrary order in a systematic way. The main difficulty
of the CEID methods, in our view, is related to the fact that it deals
directly with the electronic density matrix. As a result, certain
approximations (e.g. Hartree-Fock) for it are inevitable to facilitate
the solution of the CEID equations. 

In principle, this difficulty is circumvented in field-theoretical
methods in which electrons are treated via many-body Green's functions
(see, e.g., \cite{Mahan-book}); the Green's functions represent a
more convenient tool than the electronic (reduced) density matrix
itself involved directly in CEID. The Green's functions based techniques
have been for a long time applied to treating interacting electron-nuclear
(in fact, electron-phonon) systems at equilibrium \cite{DiVentra-book,Hedin-Lundqvist-1970,Maksimov-SovPhJETP-1976,Giustino-RMP-2017}.
To study non-equilibrium phenomena, such as, e.g., the effect of phonons
on quantum transport through a molecular junction \cite{DiVentra-book,McEniry-PRB-2008}
or carrier dynamics in semiconductors \cite{Marini-JPCM-2013}, one
has to consider non-equilibrium techniques based on non-equilibrium
Green's functions (NEGF) \cite{Keldysh-JETP-1965,Kadanoff-Baym-book,Stefanucci-Leeuwen}.
These methods are very useful and powerful in calculating, e.g., electronic
densities (occupations), currents and phonon spectra at general non-equilibrium
conditions and for a wide class of systems with either open or periodic
boundary conditions. Their main limitation, however, as far as our
main goal here is concerned, is that they can only be used in calculating
observables which are expressed via an even number of field (or creation
and annihilation) operators, such as, e.g., the electronic density
and current. However, they are not suitable for calculating atomic
trajectories as atomic positions are linear in field operators; hence,
atoms are simply assumed to oscillate around their equilibrium positions
in these methods.

In this paper we propose a general non-zero temperature method which
enables one to establish quantum ``equations of motion'' for the
expectation values of atomic positions, $\left\langle x_{A}\right\rangle _{t}$,
i.e. their mean trajectory, for arbitrary electron-nuclear system
with either periodic or open boundary conditions. The notion of the
PES is not invoked here, which we consider an advantage. In this method,
at variance with the CEID, electronic NEGF is employed instead of
the electronic reduced density matrix, which enables one to apply
this method at well known levels of approximation \cite{Stefanucci-Leeuwen}
to a wide class of non-equilibrium phenomena and virtually any system,
ranging from molecules to condensed phases and molecular junctions.
The obtained equations of motion have a stochastic form, i.e. they
contain three types of Gaussian noises which are correlated with each
other in a certain way via the electronic NEGF. Our method originates
from a few powerful ideas that were put forward a long time ago by
Hedeg{\aa}rd \cite{Hedegard-PRB-1987} and then recently extended
to current carrying molecular junctions in \cite{Brandbyge-PRB-2012},
that allowed to express the reduced density matrix of nuclei in the
coordinate representation via a partial path integral taken with respect
to the nuclear subsystem, while the electronic subsystem is presented
via an influence functional with the electronic NEGF (defined in a
slightly more general way than usual). In these papers the\emph{ partitioned
approach} for the initial density matrix (at time $t_{0}$) of the
combined system was assumed ($\widehat{\rho}(t_{0})$ is a direct
product of the density matrices of electrons and nuclei) corresponding
physically to the two subsystems being completely decoupled initially.
Also, the method used in \cite{Brandbyge-PRB-2012} to obtain equations
of motion for atoms from the path-integral representation of the nuclei
reduced density matrix was largely intuitive. The approach we propose
here is a significant generalization and extension of this method.
In detail, several important advances have been made: (i) we do not
assume that initially the electronic and nuclei subsystems were decoupled;
instead, we assume that the whole combined system was at thermal equilibrium,
so that time-dependent phenomena can in principle be considered including
the transient effects (e.g. switch on of the bias \cite{Ridley-current-PRB-2015});
(ii) the path integral method employed here is also considered as
an intermediate tool; however, the passage from the path integrals
to the equations of motion for the mean atomic positions is done rigorously,
leading to an infinite hierarchy of stochastic differential equations
for atomic positions and momenta and their various fluctuations, similarly
in spirit to the CEID equations; (iii) by employing an expansion of
the electron-nuclear interaction term around the mean atomic trajectory,
similarly in spirit to CEID and Ref. \cite{LK-c_number-PRB-2017},
which was done up to the second order, we are able to consider a general
non-equilibrium situation, whereby atoms do not merely oscillate around
their equilibrium positions, but may move along more general trajectories
(e.g. as in photo-induced dissociation reactions or during STM manipulation). 

In the coming sections we shall present the complete formulation of
the main equations of the method and the necessary detailed derivations.
No implementation and calculations with this method are yet available
and hence will not be presented here; this is left for future work.

\section{Theory}

\subsection{Hamiltonian}

At initial time $t_{0}$ the entire system (electrons and nuclei)
is assumed to be at thermal equilibrium with temperature $T$, and
described by the initial Hamiltonian 
\begin{equation}
\mathcal{H}^{0}=\mathcal{H}_{1}^{0}(x,p)+\mathcal{H}_{2}^{0}+\mathcal{H}_{12}^{0}(x)\label{eq:initial_Hamiltonian}
\end{equation}
where $\mathcal{H}_{1}^{0}$ describes the nuclear subsystem to be
considered explicitly with coordinates and momenta $x=\left\{ x_{A}\right\} $
and $p=\left\{ p_{A}\right\} $ (with $A$ designating an nuclear
degree of freedom), $\mathcal{H}_{2}^{0}$ is the Hamiltonian of the
electrons in the whole system, and $\mathcal{H}_{12}^{0}$ describes
the electron-nuclear interaction. We only show explicitly the dependence
of the Hamiltonian on the nuclear coordinates and momenta. Note that
although interaction of all nuclei with electrons is taken into account,
not all nuclei may be allowed to displace from their equilibrium positions;
only those allowed to move are explicitly included in the subsystem
(or region) 1 and hence presented in the part $\mathcal{H}_{1}^{0}(x,p)+\mathcal{H}_{12}^{0}(x)$
of the Hamiltonian and therefore included in $x$. For instance, in
the case of a molecule interacting with an external field we may consider
all its atoms to be allowed to displace, and in this case their displacements
will be considered explicitly. In the case of a molecular junction
only atoms in the central region may be considered explicitly; all
other atoms belonging to the leads will be frozen and not included
in $x$. The sum of the last two operators, $H_{2}^{0}=\mathcal{H}_{2}^{0}+\mathcal{H}_{12}^{0}(x)$,
constitutes the electrons-only Hamiltonian (for which we shall adopt
the second quantization later on). No assumptions are made concerning
the form of the nuclear-only, $\mathcal{H}_{1}^{0}$, and electron-only,
$\mathcal{H}_{2}^{0}$, Hamiltonians at this stage, they could be
as complex as required. 

The Hamiltonian of the whole system at later times $t>t_{0}$ can
be split in a similar way,

\begin{equation}
\mathcal{H}=\mathcal{H}_{1}(x,p)+\mathcal{H}_{2}+\mathcal{H}_{12}(x)\label{eq:Hamiltonian_1+2+12}
\end{equation}
and it does not need to coincide with $\mathcal{H}^{0}$, as it may
depend explicitly on time, e.g. due to an external field contained
in $\mathcal{H}_{1}+\mathcal{H}_{2}$.

The interaction between electrons and nuclei that are free to move,
$\mathcal{H}_{12}^{0}$ and $\mathcal{H}_{12}$, will be treated approximately
in the following way: we shall expand this part of the Hamiltonian
in terms of nuclear displacements. Two cases need to be considered:
(i) initial state of the whole system at time $t_{0}$ and (ii) later
times, $t>t_{0}$. In the former case nuclei from region 1 are displaced
from their \emph{equilibrium positions} $x^{0}=\left(x_{A}^{0}\right)$,
and $\mathcal{H}_{12}^{0}$ is expanded up to the second order in
terms of them: 
\begin{equation}
\mathcal{H}_{12}^{0}\equiv\sum_{nm\in C}V_{nm}^{0}\left(x\right)c_{n}^{\dagger}c_{m}\label{eq:el-ph-coupling-H_12-ini}
\end{equation}
where $n$, $m$ are indices of the localized atomic basis placed
on atoms in region 1 (this basis forms a set of orbitals $C$) to
represent electrons, $c_{n}^{\dagger}$ and $c_{m}$ are the corresponding
electronic creation and annihilation operators in that region, while 

\begin{equation}
V_{nm}^{0}(x)=\sum_{A}V_{nm}^{A,0}u_{A}+\frac{1}{2}\sum_{AB}V_{nm}^{AB,0}u_{A}u_{B}\label{eq:V_mn-init}
\end{equation}
are the corresponding matrix elements that depend on the nuclear displacements
$u_{A}=x_{A}-x_{A}^{0}$. Note that the free term in the expansion,
corresponding to zero displacements, is incorporated into the electrons-only
Hamiltonian $\mathcal{H}_{2}^{0}$. 

We need to have in mind that our goal here is to be able to study
nuclear dynamics, and hence our nuclei may not simply oscillate, but
follow a more complex trajectory at later times, $t>t_{0}$. To simplify
the problem, in the spirit of CEID \cite{McEniry-EPJB-2010} we shall
adopt a \emph{harmonization approximation} \cite{LK-c_number-PRB-2017}
in which nuclear positions are assumed to deviate no more than quadratically
from their ``exact'' instantaneous positions given by the mean nuclear
trajectory $\left\langle x\right\rangle _{t}=\left(\left\langle x_{A}\right\rangle _{t}\right)$,
where $\left\langle x_{A}\right\rangle _{t}=\text{Tr}\left[\rho(t)x_{A}\right]$
and $\rho(t)$ is the density matrix of the combined system ``electrons
+ nuclei'': 
\begin{equation}
\mathcal{H}_{12}\equiv\sum_{nm\in C}V_{nm}c_{n}^{\dagger}c_{m}\label{eq:el-ph-coupling-H_12-t_gt_t0}
\end{equation}
\begin{equation}
V_{nm}=\sum_{A}V_{nm}^{A}u_{A}+\frac{1}{2}\sum_{AB}V_{nm}^{AB}u_{A}u_{B}\label{eq:V_mn-t_gt_t0}
\end{equation}
where $u_{A}=x_{A}-\left\langle x_{A}\right\rangle _{t}$. Here the
expansion coefficients $V_{mn}^{A}$ and $V_{mn}^{AB}$ will depend
explicitly on the mean trajectory $\left\langle x\right\rangle _{t}$,
and hence, on time $t$. Similarly to the equilibrium case, Eqs. (\ref{eq:el-ph-coupling-H_12-ini})
and (\ref{eq:V_mn-init}), the free term in the expansion of $\mathcal{H}_{12}$
is incorporated into $\mathcal{H}_{2}$, so that the latter becomes
implicitly time dependent via the dependence of $V_{mn}$ on the mean
trajectory, $\mathcal{H}_{2}\equiv\mathcal{H}_{2}(t)$.

We stress here that the mean trajectory is not yet known, and our
goal in this work is to derive an appropriate equation of motion for
it. Since from the very beginning parameters of the Hamiltonian are
assumed to depend on the mean trajectory, the equation of motion we
are after may become non-linear, so that only numerical solution of
these equations is anticipated. 

\subsection{Influence functional\label{subsec:Influence-functional}}

We shall start by deriving an explicit expression for the nuclear
density matrix, reduced with respect to the electronic subsystem and
written in the nuclear coordinate representation by means of the path
integrals method. The detailed derivation of all cases needed here
is given in Appendix A.

As was shown in Ref. \cite{Hedegard-PRB-1987}, the full propagator
$\widehat{U}\left(t_{1},t_{0}\right)$ of the whole combined system,
written in the coordinate representation \emph{with respect to the
nuclear subsystem}, $\left.\left\langle x_{1}\left|\widehat{U}\left(t_{1},t_{0}\right)\right|x_{0}\right.\right\rangle $,
can be expressed via a path integral over nuclear trajectories:
\begin{equation}
\left.\left\langle x_{1}\left|\widehat{U}\left(t_{1},t_{0}\right)\right|x_{0}\right.\right\rangle =\int_{x\left(t_{0}\right)=x_{0}}^{x\left(t_{1}\right)=x_{1}}\mathcal{D}x(t)e^{iS_{1}\left[x(t)\right]/\hbar}\widehat{U}_{2}\left(t_{1},t_{0}\right)\label{eq:ion-propagator-el-operator}
\end{equation}
where 
\begin{equation}
\widehat{U}_{2}\left(t,t^{\prime}\right)=\widehat{\mathcal{T}}_{+}\exp\left\{ -\frac{i}{\hbar}\int_{t^{\prime}}^{t}\left[\mathcal{H}_{2}\left(t^{\prime\prime}\right)+\mathcal{H}_{12}\left(x\left(t^{\prime\prime}\right)\right)\right]dt^{\prime\prime}\right\} \label{eq:electronic-propagator}
\end{equation}
is the electronic propagator (and hence the subscript $2$), in which
the trajectory $x(t)$ of nuclei enters as a ``classical'' \emph{fixed}
function (and hence serves as a parameter) via the coupling term,
$\mathcal{H}_{12}$. The latter depends explicitly on time via its
dependence on the trajectory $x\left(t\right)$ in the path integral,
i.e. $\mathcal{H}_{12}$ is expanded as in Eqs. (\ref{eq:el-ph-coupling-H_12-t_gt_t0})
and (\ref{eq:V_mn-t_gt_t0}) with the displacements given by functions
$u_{A}(t)=x_{A}(t)-\left\langle x_{A}\right\rangle _{t}$. Recall
that $\mathcal{H}_{2}$ also depends on time, since nuclei are assumed
to be clumped in their average positions $\left\langle x_{A}\right\rangle _{t}$
on the mean trajectory at each time $t$. The propagator satisfies
usual equations of motion:
\[
i\hbar\partial_{t}\widehat{U}_{2}\left(t,t^{\prime}\right)=\left[\mathcal{H}_{2}\left(t\right)+\mathcal{H}_{12}\left(x\left(t\right)\right)\right]\widehat{U}_{2}\left(t,t^{\prime}\right)
\]
\begin{equation}
-i\hbar\partial_{t^{\prime}}\widehat{U}_{2}\left(t,t^{\prime}\right)=\widehat{U}_{2}\left(t,t^{\prime}\right)\left[\mathcal{H}_{2}\left(t^{\prime}\right)+\mathcal{H}_{12}\left(x\left(t^{\prime}\right)\right)\right]\label{eq:EoM-for-t-propagator}
\end{equation}
Next, $S_{1}\left[x(t)\right]$ in Eq. (\ref{eq:ion-propagator-el-operator})
is the classical action associated with the isolated nuclear subsystem
described by $\mathcal{H}_{1}$ only (which may depend on time). Finally,
$\widehat{\mathcal{T}}$ is the time-ordering operator arranging operators
in the exponent by their times ascending from right to left.

It is essential to stress, that the propagator (\ref{eq:ion-propagator-el-operator})
is an operator with respect to the electronic degrees of freedom,
but is a classical object as far as the nuclei are concerned. This
form is a hybrid between the Feynman (classical) and usual quantum
(operator) representations of the propagator.

The evolution of the total density matrix $\widehat{\rho}\left(t_{1}\right)$
for the combined system (electrons + nuclei) is given by the corresponding
solution of the Liouville equation, 
\begin{equation}
\widehat{\rho}\left(t_{1}\right)=\widehat{U}\left(t_{1},t_{0}\right)\widehat{\rho}\left(t_{0}\right)\widehat{U}\left(t_{0},t_{1}\right)\label{eq:tot-DM-from-Liouville}
\end{equation}
which in the coordinate representation, again written with respect
to the nuclei only, reads:
\[
\left.\left\langle x_{1}\left|\rho\left(t_{1}\right)\right|\right.x_{0}\right\rangle =\int dx_{2}dx_{3}\left.\left\langle x_{1}\left|\widehat{U}\left(t_{1},t_{0}\right)\right|x_{2}\right.\right\rangle \left.\left\langle x_{2}\left|\rho\left(t_{0}\right)\right|\right.x_{3}\right\rangle \left.\left\langle x_{3}\left|\widehat{U}\left(t_{0},t_{1}\right)\right|x_{0}\right.\right\rangle 
\]
\begin{equation}
=\int dx_{2}dx_{3}\int_{x\left(t_{0}\right)=x_{2}}^{x\left(t_{1}\right)=x_{1}}\mathcal{D}x(t)\,\int_{x^{\prime}\left(t_{1}\right)=x_{0}}^{x^{\prime}\left(t_{0}\right)=x_{3}}\mathcal{D}x^{\prime}(t)\,e^{i\left(S_{1}\left[x(t)\right]-S_{1}\left[x^{\prime}(t)\right]\right)/\hbar}\,\widehat{U}_{2}\left(t_{1},t_{0}\right)\left.\left\langle x_{2}\left|\rho\left(t_{0}\right)\right|\right.x_{3}\right\rangle \widehat{U}_{2}^{\dagger}\left(t_{1},t_{0}\right)\label{eq:tot-DM-via-path-ints}
\end{equation}
Here the propagator $\widehat{U}_{2}\left(t_{1},t_{0}\right)$ depends
explicitly on the nuclear trajectories $x(t)$ taken forward in time
between $t_{0}$ and $t_{1}$, while the propagator $\widehat{U}_{2}^{\dagger}\left(t_{1},t_{0}\right)=\widehat{U}_{2}\left(t_{0},t_{1}\right)$
depends explicitly on the nuclear trajectories $x^{\prime}(t)$ taken
backward in time, from $t_{1}$ to $t_{0}$. We shall explicitly indicate
this by writing $\widehat{U}_{2}\left(t_{1},t_{0}\right)_{x(t)}$
and $\widehat{U}_{2}\left(t_{0},t_{1}\right)_{x^{\prime}(t)}$ for
the two propagators.

The matrix element $\left.\left\langle x_{2}\left|\widehat{\rho}\left(t_{0}\right)\right|\right.x_{3}\right\rangle $
of the initial density matrix is still an operator for electrons,
and hence cannot be permuted with the two electronic propagators on
both sides of it. However, assuming that the whole system was at thermal
equilibrium at the initial time $t_{0}$, 
\begin{equation}
\widehat{\rho}\left(t_{0}\right)=Z_{0}^{-1}e^{-\beta\left(\mathcal{H}_{0}-\mu N\right)}\;,\quad Z_{0}=\text{Tr}\left[e^{-\beta\left(\mathcal{H}_{0}-\mu N\right)}\right]\label{eq:Rho-0}
\end{equation}
where $Z_{0}$ is the partition function of the combined system at
equilibrium at time $t_{0}$, $\beta=1/k_{B}T$ is the inverse temperature,
$\mu$ chemical potential and $N$ number operator for the electrons.
The $-\mu N$ term is convenient to absorb in the part $\mathcal{H}_{2}^{0}$
of the initial Hamiltonian $\mathcal{H}^{0}$, and this is what is
implied in what follows. Since the initial density matrix is not assumed
here as a direct product of the nuclear and electronic density matrices,
our method is partition-less.

Using a similar argument as in Ref. \cite{Hedegard-PRB-1987}, one
can write the matrix element $\left.\left\langle x_{2}\left|\rho\left(t_{0}\right)\right|\right.x_{3}\right\rangle $
via an imaginary time path integral with respect to nuclei only (i.e.
keeping it still as an operator in the electronic Hilbert space):
\begin{equation}
\left.\left\langle x_{2}\left|\widehat{\rho}\left(t_{0}\right)\right|\right.x_{3}\right\rangle =\frac{1}{Z_{0}}\int_{\overline{x}(0)=x_{3}}^{\overline{x}(\beta\hbar)=x_{2}}\mathcal{D}\overline{x}\left(\tau\right)\,e^{-S_{1}^{0}\left[\overline{x}\left(\tau\right)\right]/\hbar}\widehat{U}_{2}\left(\beta\hbar,0\right)\label{eq:initial-DM-via-path-int}
\end{equation}
Here $S_{1}^{0}\left[\overline{x}(\tau)\right]$ is the Euclidean
action associated with the initial nuclear Hamiltonian $\mathcal{H}_{1}^{0}(x,p)$,
and 
\begin{equation}
\widehat{U}_{2}\left(\tau,\tau^{\prime}\right)=\overline{\mathcal{T}}\,\exp\left\{ -\frac{1}{\hbar}\int_{\tau^{\prime}}^{\tau}\left[\mathcal{H}_{2}^{0}+\mathcal{H}_{12}\left(\overline{x}(\tau^{\prime\prime})\right)\right]d\tau^{\prime\prime}\right\} \label{eq:imag-time-propag-operator}
\end{equation}
is the Euclidean propagation operator and $\overline{\mathcal{T}}$
the imaginary time ordering operator. In particular, $\widehat{U}_{2}\left(\beta\hbar,0\right)$
evolves the electronic subsystem in the imaginary time from $\tau^{\prime}=0$
to $\tau=\beta\hbar$. Here yet again the trajectory $\overline{x}(\tau)$
of nuclei (entering via the coupling term) is fixed, so that the propagation
operator explicitly depends on it, to be indicated as $\widehat{U}_{2}\left(\beta\hbar,0\right)_{\overline{x}(\tau)}$. 

The operator (\ref{eq:imag-time-propag-operator}) satisfies 
\[
-\hbar\partial_{\tau}\widehat{U}_{2}\left(\tau,\tau^{\prime}\right)=\left[\mathcal{H}_{2}^{0}+\mathcal{H}_{12}\left(\overline{x}(\tau)\right)\right]\widehat{U}_{2}\left(\tau,\tau^{\prime}\right)
\]
\begin{equation}
\hbar\partial_{\tau^{\prime}}\widehat{U}_{2}\left(\tau,\tau^{\prime}\right)=\widehat{U}_{2}\left(\tau,\tau^{\prime}\right)\left[\mathcal{H}_{2}^{0}+\mathcal{H}_{12}\left(\overline{x}(\tau^{\prime})\right)\right]\label{eq:EoM-for-tau-evol-operator}
\end{equation}

Using Eqs. (\ref{eq:tot-DM-via-path-ints}) and (\ref{eq:initial-DM-via-path-int}),
we can write for the density matrix of the combined system, still
in the coordinate representation for nuclei, an expression: 
\[
\left.\left\langle x_{1}\left|\widehat{\rho}\left(t_{1}\right)\right|\right.x_{0}\right\rangle =\frac{1}{Z_{0}}\int dx_{2}dx_{3}\int\mathcal{D}x(t)\,\int\mathcal{D}x^{\prime}(t)\,\int\mathcal{D}\overline{x}\left(\tau\right)\,e^{\frac{i}{\hbar}\left(S_{1}\left[x(t)\right]-S_{1}\left[x^{\prime}(t)\right]\right)-\frac{1}{\hbar}S_{1}^{0}\left[\overline{x}\left(\tau\right)\right]}
\]
\[
\times\,\widehat{U}_{2}\left(t_{1},t_{0}\right)_{x(t)}\widehat{U}_{2}\left(\beta\hbar,0\right)_{\overline{x}(\tau)}\widehat{U}_{2}^{\dagger}\left(t_{1},t_{0}\right)_{x^{\prime}(t)}
\]
The obtained expression is still an operator for the electronic subsystem.
To obtain the reduced density matrix for the nuclear subsystem we
are interested in, we have to take a trace (to be denoted $\text{Tr}_{2}\left(\ldots\right)$)
with respect to the Hilbert space associated with the electrons: 
\[
\left.\left\langle x_{1}\left|\widehat{\rho}_{ions}\left(t_{1}\right)\right|\right.x_{0}\right\rangle =\frac{1}{Z_{0}}\int dx_{2}dx_{3}\int\mathcal{D}x(t)\,\int\mathcal{D}x^{\prime}(t)\,\int\mathcal{D}\overline{x}\left(\tau\right)\,e^{\frac{i}{\hbar}\left(S_{1}\left[x(t)\right]-S_{1}\left[x^{\prime}(t)\right]\right)-\frac{1}{\hbar}S_{1}^{0}\left[\overline{x}\left(\tau\right)\right]}
\]
\begin{equation}
\times\text{Tr}_{2}\left[\widehat{U}_{2}\left(\beta\hbar,0\right)_{\overline{x}(\tau)}\widehat{U}_{2}\left(t_{0},t_{1}\right)_{x^{\prime}(t)}\widehat{U}_{2}\left(t_{1},t_{0}\right)_{x(t)}\right]\label{eq:red-DM-of-ions}
\end{equation}
where the cyclic invariance of the trace has been used. This expression
is not an operator anymore.

\begin{figure}
\begin{centering}
\includegraphics[height=4cm]{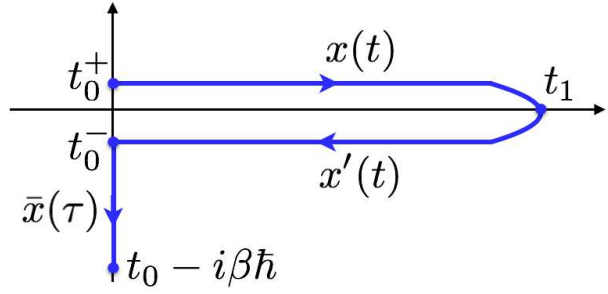}
\par\end{centering}
\caption{Konstantinov-Perel's contour $\gamma$. On the horizontal upper (forward)
track of the contour $\gamma$ the coupling part of the Hamiltonian
depends explicitly on the fixed nuclear trajectory $x(t)$, the horizontal
lower (backward) track - on the fixed nuclear trajectory $x^{\prime}(t)$,
while the vertical track - on the fixed nuclear trajectory $\overline{x}(\tau)$
corresponding to Euclidean evolution. \label{fig:Konstantinov-Perel-contour}}
\end{figure}

The product of the three operators under the trace, if read from right
to left, has first a forward propagation from $t_{0}$ to $t_{1}$,
then a backward propagation from $t_{1}$ to $t_{0}$, and, finally,
imaginary time propagation from $t_{0}^{-}\equiv t_{0}-i0$ to $t_{0}-i\beta\hbar$.
It is convenient to introduce a single contour consisting of these
three parts: $t_{0}^{+}\equiv t_{0}+i0\rightarrow t_{1}\rightarrow t_{0}^{-}\rightarrow t_{0}-i\beta\hbar$.
This is known as the Konstantinov-Perel's contour, Fig. \ref{fig:Konstantinov-Perel-contour},
which will be denoted hereafter as $\gamma$. It is essential that
the time $t_{1}$ on $\gamma$ is \emph{fixed} as corresponding to
the observation time (see the left hand side of Eq. (\ref{eq:red-DM-of-ions})).
Also, it is essential to remember that on each of the three parts
(tracks) of the contour the coupling Hamiltonian $\mathcal{H}_{12}$
is taken with a particular fixed nuclear trajectory, either $x(t)$,
or $x^{\prime}(t)$ on the horizontal tracks, and $\mathcal{H}_{12}^{0}$
with $\overline{x}(\tau)$ on the vertical one. This brings a time
dependence to the electronic problem; it is additional to any other
existing, e.g. due to the time dependent field and the harmonisation
approximation. 

Hence, denoting the product as a single operator,
\begin{equation}
\widehat{U}_{\gamma}\left(t_{0}-i\beta\hbar,t_{0}^{+}\right)=\widehat{U}_{2}\left(\beta\hbar,0\right)_{\overline{x}(\tau)}\widehat{U}_{2}\left(t_{0},t_{1}\right)_{x^{\prime}(t)}\widehat{U}_{2}\left(t_{1},t_{0}\right)_{x(t)}\label{eq:tot-evolution-operator}
\end{equation}
and using the fact that the order in which the operators appear in
the product is fixed, it is possible to rewrite it as a single evolution
operator over the whole contour: 
\begin{equation}
\widehat{U}_{\gamma}\left(t_{0}-i\beta\hbar,t_{0}^{+}\right)=\widehat{\mathcal{T}}_{\gamma}\exp\left\{ -\frac{i}{\hbar}\int_{\gamma}H_{2}\left(z_{1}\right)dz_{1}\right\} \label{eq:tot-evolution-operator-over-gamma}
\end{equation}
Here $\widehat{\mathcal{T}}_{\gamma}$ is the time ordering operator
on the contour with the direction of $z_{1}\in\gamma$ increasing
as shown by arrows in Fig. \ref{fig:Konstantinov-Perel-contour}.
The integral over the contour consists of a sum of three integrals:
over the upper, then lower and finally over the vertical tracks with
the electronic Hamiltonian in each part of the contour (written by
the corresponding Roman letter) defined as follows:
\begin{equation}
H_{2}(z)\equiv\left\{ \begin{array}{cc}
\mathcal{H}_{2}(t)+\mathcal{H}_{12}\left(x(t)\right) & \text{if }z\in\text{upper track}\\
\mathcal{H}_{2}(t)+\mathcal{H}_{12}\left(x^{\prime}(t)\right) & \text{if }z\in\text{lower track}\\
\mathcal{H}_{2}^{0}+\mathcal{H}_{12}^{0}\left(\overline{x}(t)\right) & \text{if }z\in\text{vertical track}
\end{array}\right.\label{eq:electronic-H}
\end{equation}
We have explicitly indicated here how the electron-nuclear part depends
on the nuclear positions on each track. The electron-nuclear part
of the Hamiltonian, $\mathcal{H}_{12}$ and $\mathcal{H}_{12}^{0}$,
are given by Eqs. (\ref{eq:el-ph-coupling-H_12-ini})-(\ref{eq:V_mn-t_gt_t0}),
where atomic displacements $u_{A}$ are given somewhat differently
depending on the track on the contour $\gamma$: on the horizontal
tracks the ``classical'' displacement of the degree of freedom $A$
is defined as $u_{A}(t)=x_{A}(t)-\left\langle x_{A}\right\rangle _{t}$
(upper) and $u_{A}^{\prime}(t)=x_{A}^{\prime}(t)-\left\langle x_{A}\right\rangle _{t}$
(lower), while on the vertical track the displacement $\overline{u}_{A}\left(\tau\right)=x_{A}\left(\tau\right)-x_{A}^{0}$
is used that is defined with respect to the equilibrium positions
of the atoms. The matrices $\mathbf{V}^{A}(z)=\left(V_{nm}^{A}(z)\right)$
and $\mathbf{V}^{AB}(z)=\left(V_{nm}^{AB}(z)\right)$, see Eqs. (\ref{eq:V_mn-init})
and (\ref{eq:V_mn-t_gt_t0}), may still depend on $z\in\gamma$ on
the horizontal tracks via its possible dependence on the averages
$\left\langle x_{A}\right\rangle _{t}$, i.e. they are $\mathbf{V}^{A}(t)$
and $\mathbf{V}^{AB}(t)$ on both tracks, while on the vertical track
there is no time dependence, i.e. $\mathbf{V}^{A}\left(\tau\right)\equiv\mathbf{V}_{0}^{A}$
and $\mathbf{V}^{AB}\left(\tau\right)\equiv\mathbf{V}_{0}^{AB}$,
the values at thermal equilibrium.

In all these cases the nuclear trajectories are fixed (by the corresponding
path integrals in Eq. (\ref{eq:red-DM-of-ions})), i.e. they serve
as ``external'' parameters (functions). As will be clear later on,
the fact that the Hamiltonian on each track is different and depends
on time in this rather general way creates additional complications
in developing theory.

For the following it is convenient to introduce a more general evolution
operator between any two variables $z$ and $z^{\prime}$ somewhere
on $\gamma$,
\begin{equation}
\widehat{U}_{\gamma}\left(z,z^{\prime}\right)=\widehat{\mathcal{T}}_{\gamma}\exp\left\{ -\frac{i}{\hbar}\int_{z^{\prime}}^{z}H_{2}\left(z_{1}\right)dz_{1}\right\} \label{eq:average-of-tot-evolution-operator}
\end{equation}
The operator (\ref{eq:tot-evolution-operator-over-gamma}) is obtained
by taking $z=t_{0}-i\beta\hbar$ and $z^{\prime}=t_{0}^{+}$, and
the electronic Hamiltonian $H_{2}\left(z\right)$ on $\gamma$ is
defined as described above. By its definition (\ref{eq:average-of-tot-evolution-operator}),
$\widehat{U}_{\gamma}\left(z,z^{\prime}\right)$ is essentially a
product of the required number of evolution operators, which are necessary
to bridge the two ``times'' $z$ and $z^{\prime}$. Hence, using
Eqs. (\ref{eq:EoM-for-t-propagator}) and (\ref{eq:EoM-for-tau-evol-operator})
and the fact that on the imaginary (vertical) track $z=t_{0}-i\tau$,
the propagator $\widehat{U}_{\gamma}\left(z,z^{\prime}\right)$ is
seen to satisfy the usual equations of motion:
\[
i\hbar\partial_{z}\widehat{U}_{\gamma}\left(z,z^{\prime}\right)=H_{2}(z)\widehat{U}_{\gamma}\left(z,z^{\prime}\right)
\]
\begin{equation}
-i\hbar\partial_{z^{\prime}}\widehat{U}_{\gamma}\left(z,z^{\prime}\right)=\widehat{U}_{\gamma}\left(z,z^{\prime}\right)H_{2}\left(z^{\prime}\right)\label{eq:EoM-for-tot-propagator}
\end{equation}

Concluding, the trace of the product of three electronic propagators
can be written as 
\begin{equation}
\text{Tr}_{2}\left[\widehat{U}_{2}\left(\beta\hbar,0\right)_{\overline{x}(\tau)}\widehat{U}_{2}\left(t_{0},t_{1}\right)_{x^{\prime}(t)}\widehat{U}_{2}\left(t_{1},t_{0}\right)_{x(t)}\right]=\text{Tr}_{2}\left[\widehat{U}_{\gamma}\left(t_{0}-i\beta\hbar,t_{0}^{+}\right)\right]\equiv\left\langle \widehat{U}_{\gamma}\left(t_{0}-i\beta\hbar,t_{0}^{+}\right)\right\rangle _{2}\label{eq:universal-propagator-def}
\end{equation}
and hence we have to develop methods of calculating the trace of the
electronic propagation operator in the right hand side.

Following the steps of Ref. \cite{Hedegard-PRB-1987}, one can derive
a useful formula for the required expectation value $\left\langle \widehat{U}_{\gamma}\left(t_{0}-i\beta\hbar,t_{0}^{+}\right)\right\rangle _{2}$
of the evolution operator (\ref{eq:tot-evolution-operator-over-gamma})
on $\gamma$ that would enable us to define the (generalised) electronic
Green's function later on. We first introduce a parameter $\lambda$
that stipulates the strength of the coupling term: 
\begin{equation}
H_{2}^{\lambda}(z)\equiv\left\{ \begin{array}{cc}
\mathcal{H}_{2}(t)+\lambda\mathcal{H}_{12}\left(x(t)\right) & \text{if }z\in\text{upper track}\\
\mathcal{H}_{2}(t)+\lambda\mathcal{H}_{12}\left(x^{\prime}(t)\right) & \text{if }z\in\text{lower track}\\
\mathcal{H}_{2}^{0}+\lambda\mathcal{H}_{12}\left(\overline{x}(t)\right) & \text{if }z\in\text{vertical track}
\end{array}\right.\label{eq:electronic-H-lambda}
\end{equation}
Here, $H_{2}^{\lambda}(z)$ is defined with either $x(t)$, $x^{\prime}(t)$
or $\overline{x}(\tau)$, depending on the position of the variable
$z$ on the contour $\gamma$. At $\lambda=1$ we have our original
Hamiltonian (\ref{eq:electronic-H}). At $\lambda=0$ the electron-ion
coupling is completely switched off, although electrons still interact
(via $\mathcal{H}_{2}$ or $\mathcal{H}_{2}^{0}$) with nuclei clumped
at their respective positions. Correspondingly, in $\mathcal{H}_{2}$
nuclei are assumed to be at positions $\left\langle x_{A}\right\rangle _{t}$
for $z$ anywhere on the horizontal tracks, while on the vertical
track they are at equilibrium positions $x_{A}^{0}$.

Next, we define a new evolution operator via
\begin{equation}
\widehat{U}_{\gamma}^{\lambda}\left(z,z^{\prime}\right)=\widehat{\mathcal{T}}_{\gamma}\exp\left\{ -\frac{i}{\hbar}\int_{z^{\prime}}^{z}H_{2}^{\lambda}\left(z_{1}\right)dz_{1}\right\} \label{eq:evolution-operator-z-z'}
\end{equation}
between any two points $z$ and $z^{\prime}$ on $\gamma$. At $\lambda=1$
this operator goes over into the one we introduced above, Eq. (\ref{eq:average-of-tot-evolution-operator}),
and which we actually need. The new operator satisfies the usual equations
of motion:
\[
i\hbar\partial_{z}\widehat{U}_{\gamma}^{\lambda}\left(z,z^{\prime}\right)=H_{2}^{\lambda}(z)\widehat{U}_{\gamma}^{\lambda}\left(z,z^{\prime}\right)
\]
\begin{equation}
-i\hbar\partial_{z^{\prime}}\widehat{U}_{\gamma}^{\lambda}\left(z,z^{\prime}\right)=\widehat{U}_{\gamma}^{\lambda}\left(z,z^{\prime}\right)H_{2}^{\lambda}\left(z^{\prime}\right)\label{eq:EoM-lambda-evoluiton}
\end{equation}

Then, it follows from a well-known expression valid on the contour
$\gamma$ that 
\begin{equation}
\frac{\partial}{\partial\lambda}\widehat{U}_{\gamma}^{\lambda}\left(z,z^{\prime}\right)=-\frac{i}{\hbar}\int_{\gamma}dz_{1}\widehat{U}_{\gamma}^{\lambda}\left(z,z_{1}\right)\frac{\partial H_{2}^{\lambda}\left(z_{1}\right)}{\partial\lambda}\widehat{U}_{\gamma}^{\lambda}\left(z_{1},z^{\prime}\right)=-\frac{i}{\hbar}\int_{\gamma}dz_{1}\widehat{U}_{\gamma}^{\lambda}\left(z,z_{1}\right)\mathcal{H}_{12}\left(z_{1}\right)\widehat{U}_{\gamma}^{\lambda}\left(z_{1},z^{\prime}\right)\label{eq:useful-identity-for-evolution-deriv}
\end{equation}
Therefore, the derivative of the trace of the evolution operator with
the times on $\gamma$ as appearing in Eq. (\ref{eq:universal-propagator-def})
can be written as
\[
\frac{\partial}{\partial\lambda}\left\langle \widehat{U}_{\gamma}^{\lambda}\left(t_{0}-i\beta\hbar,t_{0}^{+}\right)\right\rangle _{2}=-\frac{i}{\hbar}\int_{\gamma}dz_{1}\left\langle \widehat{U}_{\gamma}^{\lambda}\left(t_{0}-i\beta\hbar,z_{1}\right)\mathcal{H}_{12}\left(z_{1}\right)\widehat{U}_{\gamma}^{\lambda}\left(z_{1},t_{0}^{+}\right)\right\rangle _{2}
\]
Dividing both sides of this equation by $\left\langle \widehat{U}_{\gamma}^{\lambda}\left(t_{0}-i\beta\hbar,t_{0}^{+}\right)\right\rangle _{2}$
and integrating with respect to $\lambda$ between $0$ and $1$,
one obtains:
\begin{equation}
\left\langle \widehat{U}_{\gamma}\left(t_{0}-i\beta\hbar,t_{0}^{+}\right)\right\rangle _{2}\equiv\left\langle \widehat{U}_{\gamma}^{\lambda}\left(t_{0}-i\beta\hbar,t_{0}^{+}\right)\right\rangle _{2}^{\lambda=1}=Z_{2}^{0}\exp\left\{ \frac{i}{\hbar}\Delta S_{eff}\left[x(t),x^{\prime}(t),\overline{x}(\tau)\right]\right\} \label{eq:average-of-evolution-via_S_eff}
\end{equation}
where we have introduced an effective action
\begin{equation}
\Delta S_{eff}\left[x(t),x^{\prime}(t),\overline{x}(\tau)\right]=-\int_{0}^{1}d\lambda\int_{\gamma}dz_{1}\,\mathcal{F}_{\lambda}\left(z_{1}\right)\label{eq:eff-action}
\end{equation}
where
\begin{equation}
\mathcal{F}_{\lambda}(z)=\frac{\left\langle \widehat{U}_{\gamma}^{\lambda}\left(t_{0}-i\beta\hbar,z\right)\mathcal{H}_{12}\left(z\right)\widehat{U}_{\gamma}^{\lambda}\left(z,t_{0}^{+}\right)\right\rangle _{2}}{\left\langle \widehat{U}_{\gamma}^{\lambda}\left(t_{0}-i\beta\hbar,t_{0}^{+}\right)\right\rangle _{2}}\label{eq:F_lambda-function}
\end{equation}
In deriving the above result, use has been made of the fact that at
$\lambda=0$ the coupling between the nuclei and electrons disappears,
and hence 
\[
\left\langle \widehat{U}_{\gamma}^{\lambda}\left(t_{0}-i\beta\hbar,t_{0}^{+}\right)\right\rangle _{2}^{\lambda=0}=\text{Tr}\left[\widehat{U}_{2}\left(t_{0}-i\beta\hbar,t_{0}^{-}\right)\widehat{U}_{2}\left(t_{0}^{-},t_{1}\right)\widehat{U}_{2}\left(t_{1},t_{0}^{+}\right)\right]_{\lambda=0}
\]
\[
=\text{Tr}\left[\widehat{U}_{2}\left(t_{0}-i\beta\hbar,t_{0}^{-}\right)\right]=\text{Tr}\left[e^{-\beta\mathcal{H}_{2}^{0}}\right]=Z_{2}^{0}
\]
is the electron-only partition function at $t_{0}$ calculated while
nuclei are clumped at their equilibrium positions $x^{0}=\left(x_{A}^{0}\right)$. 

Hence, finally, we obtain for the reduced density matrix of nuclei
an expression:
\begin{equation}
\left.\left\langle x_{1}\left|\widehat{\rho}_{ions}\left(t_{1}\right)\right|\right.x_{0}\right\rangle =\frac{Z_{2}^{0}}{Z_{0}}\int dx_{2}dx_{3}\int\mathcal{D}x(t)\,\int\mathcal{D}x^{\prime}(t)\,\int\mathcal{D}\overline{x}\left(\tau\right)\,e^{\frac{i}{\hbar}\left(S_{1}-S_{1}^{\prime}+\Delta S_{eff}\right)-\frac{1}{\hbar}S_{1}^{0}}\label{eq:final-path-int-expr-for-DM}
\end{equation}
where $S_{1}\equiv S_{1}\left[x(t)\right]$, $S_{1}^{\prime}\equiv S_{1}\left[x^{\prime}(t)\right]$,
$S_{1}^{0}\equiv S_{1}^{0}\left[\overline{x}\left(\tau\right)\right]$
and $\Delta S_{eff}$ depends on all three trajectories. The pre-factor
$Z_{2}^{0}/Z_{0}$ is a constant, which depends on temperature. We
will have to consider it later on. 

The above expression contains all the information about the electronic
subsystem in the form of the influence functional, i.e. in the expression
in the exponent, which is basically the effective action $\Delta S_{eff}$.
The obtained result generalizes the formula obtained in Refs. \cite{Hedegard-PRB-1987,Brandbyge-PRB-2012}
for the partitioned case to the one in which the electrons and nuclei
are considered coupled from the very beginning, i.e. at initial thermalization.
Hence, our method is strictly partition-less, and hence, at least
in principle, consideration of a response of the system in real time
to external time-dependent perturbations should be possible (e.g.
bias switch on). 

\subsection{Green's function\label{subsec:Green's-function}}

Because the Hamiltonian $H_{2}^{\lambda}(z)$ is different on both
horizontal tracks of $\gamma$, the two possible positions of the
initial time $t_{0}$, either on the upper (as $t_{0}^{+}$) or lower
($t_{0}^{-}$) tracks, are not equivalent. Hence, the definition of
the Heisenberg picture is not unique. Therefore, the electronic Green's
function (GF) cannot be introduced in the usual way via the operators
in the Heisenberg representation as this appears to be ambiguous.
Following Ref. \cite{Brandbyge-PRB-1995}, we generalize the definition
of the GF as follows:
\begin{equation}
G_{ab}^{\lambda}\left(z,z^{\prime}\right)=-\frac{i}{Z^{\lambda}\hbar}\left\{ \begin{array}{ccc}
\left\langle \widehat{U}_{\gamma}^{\lambda}\left(t_{0}-i\beta\hbar,z\right)c_{a}\widehat{U}_{\gamma}^{\lambda}\left(z,z^{\prime}\right)c_{b}^{\dagger}\widehat{U}_{\gamma}^{\lambda}\left(z^{\prime},t_{0}^{+}\right)\right\rangle _{2} & \text{if} & z>z^{\prime}\\
-\left\langle \widehat{U}_{\gamma}^{\lambda}\left(t_{0}-i\beta\hbar,z^{\prime}\right)c_{b}^{\dagger}\widehat{U}_{\gamma}^{\lambda}\left(z^{\prime},z\right)c_{a}\widehat{U}_{\gamma}^{\lambda}\left(z,t_{0}^{+}\right)\right\rangle _{2} & \text{if} & z<z^{\prime}
\end{array}\right.\label{eq:GF_def}
\end{equation}
where 
\begin{equation}
Z^{\lambda}=\left\langle \widehat{U}_{\gamma}^{\lambda}\left(t_{0}-i\beta\hbar,t_{0}^{+}\right)\right\rangle _{2}\label{eq:def_of_Z}
\end{equation}
and the arguments $z$ and $z^{\prime}$ in the GF could be anywhere
on the contour $\gamma$. Here, and in the following, the indices
like $a$, $b$ correspond to any atomic orbital in the whole electronic
system, either included in the set $C$ or not. We shall use indices
like $n,$ $m$ to indicate orbitals from region $C$ only. Note that
the usual definition of the GF can also be brought into the form above
where the two propagation operators between $c_{a}$ and $c_{b}^{\dagger}$
are combined into one; hence, the defined above GF satisfies usual
equations of motion. 

Note in passing an important point, which shall be used extensively
in the following, that it is e.g. insufficient to indicate whether
the GF is ``lesser'' or ``greater'' as it may also be necessary
to indicate explicitly on which tracks of $\gamma$ the two ``time''
variables actually are. This will be indicated explicitly by the indices
$+$, $-$ or $M$ depending on whether each variable is on the upper,
lower or vertical track. For instance, if both variables $z,z^{\prime}\in+$,
then we shall us the notation $\mathbf{G}_{++}\left(z,z^{\prime}\right)$,
and then one can also introduce a lesser, greater, retarded and advanced
components; if $z\in+$ and $z^{\prime}\in M$, then we shall use
$\mathbf{G}_{+M}\left(z,z^{\prime}\right)$ (which is a lesser component)
as the usual notation $\mathbf{G}^{\rceil}\left(z,z^{\prime}\right)$
is insufficient as it is still not clear on which horizontal track
the variable $z$ lies. Only for GFs defined in such a way that the
same Hamiltonian is used on both horizontal tracks, usual notations
(without $+$ and $-$) are applicable.

With this definition, the function $\mathcal{F}_{\lambda}(z)$ in
Eq. (\ref{eq:F_lambda-function}), needed for calculating the effective
action, Eq. (\ref{eq:eff-action}), becomes:
\begin{equation}
\mathcal{F}_{\lambda}(z)=-i\hbar\sum_{nm\in C}V_{nm}\left(x(z)\right)G_{mn}^{\lambda}\left(z,z^{+}\right)=-i\hbar\,\text{tr}_{C}\left\{ \mathbf{V}\left(x(z)\right)\mathbf{G}^{\lambda}\left(z,z^{+}\right)\right\} \label{eq:F_lambda-via-GF}
\end{equation}
where the trace written above using small letters corresponds to the
usual trace of a matrix; $\mathbf{V}$ and $\mathbf{G}$ are the matrices
composed of the matrix elements $V_{nm}$ of the electron-nuclear
coupling and $G_{mn}$ of the GF. Hence, the effective action reads
\begin{equation}
\Delta S_{eff}\left[x(t),x^{\prime}(t),\overline{x}(\tau)\right]=i\hbar\int_{0}^{1}d\lambda\int_{\gamma}dz\,\text{tr}_{C}\left\{ \mathbf{V}\left(x(z)\right)\mathbf{G}^{\lambda}\left(z,z^{+}\right)\right\} \equiv i\hbar\int_{0}^{1}d\lambda\int_{\gamma}dz\,\text{tr}_{C}\left\{ \mathbf{V}(x\left(z\right))\mathbf{G}^{\lambda,<}\left(z,z\right)\right\} \label{eq:eff-action-via-GF}
\end{equation}
In the last two equations $z^{+}$ is infinitesimally ``later''
on the contour $\gamma$ than $z$, so that essentially only the lesser
GF is needed on the contour (the last equality). This formula is analogous
to the one derived initially in Ref. \cite{Brandbyge-PRB-2012,Brandbyge-PRB-1995}
based on the partitioned approach; as we shall see, going beyond this
approximation simply extends the integration from Keldysh to Konstantinov-Perel
contour. 

Writing the contour integral in Eq. (\ref{eq:eff-action-via-GF})
explicitly, we obtain for the effective action:
\[
\Delta S_{eff}\left[x(t),x^{\prime}(t),\overline{x}(\tau)\right]
\]
\begin{equation}
=i\hbar\int_{0}^{1}d\lambda\left\{ \int_{t_{0}}^{t_{1}}dt\,\text{tr}_{C}\left[\mathbf{V}_{+}\left(t\right)\mathbf{G}_{++}^{\lambda,<}\left(t,t\right)-\mathbf{V}_{-}\left(t\right)\mathbf{G}_{--}^{\lambda,<}\left(t,t\right)\right]-i\int_{0}^{\beta\hbar}d\tau\,\text{tr}_{C}\left[\mathbf{V}_{M}\left(\tau\right)\mathbf{G}_{MM}^{\lambda,<}\left(\tau,\tau\right)\right]\right\} \label{eq:eff-action-via-ful-GF}
\end{equation}

The electronic GF introduced above depends explicitly on the strength
parameter $\lambda$. At $\lambda=0$, no electron-nuclear interaction
exists and nuclei are assumed to be clumped at either positions $\left\langle x_{A}\right\rangle _{t}$
or $x_{A}^{0}$, as explained above. The GF, corresponding to the
absence of interaction between electrons and nuclear displacements,
will be denoted $\mathbb{G}\left(z,z^{\prime}\right)=\left(\mathbb{G}_{ab}\left(z,z^{\prime}\right)\right)$.
This GF corresponds to the Hamiltonian $H_{2}^{\lambda=0}$, Eq. (\ref{eq:electronic-H-lambda}),
where electron-nuclear interaction is included only in the zeroth
order with nuclei following the mean trajectory $\left\langle x\right\rangle _{t}$
on both horizontal tracks and placed at equilibrium $x^{0}$ on the
vertical track. Correspondingly, the Hamiltonian on the horizontal
tracks is the same for $\mathbb{G}$ (and the usual Langreth rules
\cite{Stefanucci-Leeuwen} apply) and hence this particular GF is
exactly the same as the usual one; there is no need to indicate explicitly
on which tracks the times are, one can use lesser, greater, etc. (usual)
notations for $\mathbb{G}$ without confusion.

The interactions of electrons with nuclear displacements, Eqs. (\ref{eq:el-ph-coupling-H_12-ini})
and (\ref{eq:el-ph-coupling-H_12-t_gt_t0}), is a one-particle operator
\[
\lambda\mathcal{H}_{12}(z)=\lambda\sum_{nm\in C}V_{nm}(z)c_{n}^{\dagger}c_{m}
\]
(and similarly for $\mathcal{H}_{12}^{0}$), and hence the two GF,
$\mathbb{G}$ and $\mathbf{G}^{\lambda}$, are related via the Dyson
equation: 
\begin{equation}
\mathbf{G}^{\lambda}\left(z,z^{\prime}\right)-\mathbb{G}\left(z,z^{\prime}\right)=\lambda\int_{\gamma}dz_{1}\,\mathbb{G}\left(z,z_{1}\right)\mathbf{V}\left(z_{1}\right)\mathbf{G}^{\lambda}\left(z_{1},z^{\prime}\right)=\lambda\int_{\gamma}dz_{1}\,\mathbf{G}^{\lambda}\left(z,z_{1}\right)\mathbf{V}\left(z_{1}\right)\mathbb{G}\left(z_{1},z^{\prime}\right)\label{eq:T1-Dayson-eq}
\end{equation}
Note that in the following only this equation will be explored. Since
it is valid for any electronic Hamiltonian that may even include electronic
correlations, the method to be adopted below is general; we shall
assume, however, that the calculation of the unperturbed GF $\mathbb{G}\left(z,z^{\prime}\right)$
is feasible. Note that this equation is also valid for the $CC$ block
of the GF, $\mathbf{G}_{CC}^{\lambda}\left(z,z^{\prime}\right)$,
in the situation when only within region $C$ nuclei are allowed to
move (see Appendix B); in this case the vector function $\mathbf{V}(z)\rightarrow\mathbf{V}_{C}(z)$
is defined only on this region's orbitals ($V_{ab}\neq0$ only if
the orbitals $a,b\in C$). 

Further, we note that since the matrices $\mathbf{V}_{\pm}(t)$ and
$\mathbf{V}_{M}(\tau)$ are non-zero only within region $C$, only
$CC$ elements of the GF are needed in Eq. (\ref{eq:eff-action-via-ful-GF})
to calculate the effective action (the influence functional). This
also means that only $CC$ elements of the unperturbed GF $\mathbb{G}\left(z,z^{\prime}\right)$
are needed in the Dyson equation. If the set $C$ does not cover the
whole system, e.g. it corresponds only to the central region (molecule)
in the quantum transport setup, then calculation of the contribution
due to other orbitals (of the leads in the case of the quantum transport)
is required. If the interaction between orbitals in $C$ and the rest
of the system is described in the one-electron approximation, then
the contribution to the $CC$ block, $\mathbb{G}_{CC}\left(z,z^{\prime}\right)$,
from the rest of the system appears in the usual way via a self-energy
(Appendix B). 

Hence, the calculation of the action, Eq. (\ref{eq:eff-action-via-ful-GF}),
requires three lesser components of the total electronic Green's function
with the coupling added using the strength $\lambda$. This requires
solving the Dyson equation (\ref{eq:T1-Dayson-eq}). We shall do it
approximately by noting that the atomic displacements $u_{A}(t)$,
$u_{A}^{\prime}(t)$ and $\overline{u}_{A}\left(\tau\right)$, with
respect to which the paths integrals are actually calculated in Eq.
(\ref{eq:final-path-int-expr-for-DM}), enter here only via the matrix
$\mathbf{V}(z)=\left(V_{nm}\left(x(z)\right)\right)$. We shall expand
the GF in powers of this matrix reiterating the Dyson equation. Since
the effective action (\ref{eq:eff-action-via-GF}) is already proportional
to such a matrix, and a progress can only be made if the effective
action is quadratic with respect to atomic displacements, we shall
limit ourselves with the first order term only (cf. \cite{Brandbyge-PRB-2012}):
\begin{equation}
\mathbf{G}^{\lambda}\left(z,z^{\prime}\right)\simeq\mathbb{G}\left(z,z^{\prime}\right)+\lambda\int_{\gamma}dz_{1}\mathbb{G}\left(z,z_{1}\right)\mathbf{V}\left(z_{1}\right)\mathbb{G}\left(z_{1},z^{\prime}\right)\label{eq:GF-to-the-1st-order}
\end{equation}
We stress, that this approximation does not imply that the electron-nuclear
coupling is small; rather, we treat nuclear displacements from the
corresponding mean positions (either $\left\langle x_{A}\right\rangle _{t}$
or $x_{A}^{0}$ on the horizontal and vertical tracks, respectively)
as small. Calculation of the contour integral in the above equation
requires an appropriate generalization of the Langreth rules which
the reader can find in Appendix C.

We now have everything we need to calculate the effective action. 

\subsection{Calculation of the effective action}

Using the generalised Langreth rules in Eq. (\ref{eq:GF-to-the-1st-order}),
one can calculate all three components of the lesser GF required in
Eq. (\ref{eq:eff-action-via-ful-GF}): on the upper track,
\[
\mathbf{G}_{++}^{\lambda<}\left(t,t\right)\simeq\mathbb{G}^{<}\left(t,t\right)-i\lambda\int_{0}^{\beta\hbar}d\tau\,\mathbb{G}^{\rceil}\left(t,\tau\right)\mathbf{V}_{M}\left(\tau\right)\mathbb{G}^{\lceil}\left(\tau,t\right)+\lambda\int_{t_{0}}^{t_{1}}dt^{\prime}\,\left[\mathbb{G}^{r}\left(t,t^{\prime}\right)\mathbf{V}_{+}\left(t^{\prime}\right)\mathbb{G}^{<}\left(t^{\prime},t\right)\right.
\]
\begin{equation}
\left.+\mathbb{G}^{<}\left(t,t^{\prime}\right)\mathbf{V}_{+}\left(t^{\prime}\right)\mathbb{G}^{a}\left(t^{\prime},t\right)+\mathbb{G}^{<}\left(t,t^{\prime}\right)\left(\mathbf{V}_{+}\left(t^{\prime}\right)-\mathbf{V}_{-}\left(t^{\prime}\right)\right)\mathbb{G}^{>}\left(t^{\prime},t\right)\right]\label{eq:G++_lesser}
\end{equation}
on the lower track,
\begin{equation}
\mathbf{G}_{--}^{\lambda<}\left(t,t\right)\simeq\mathbb{G}^{<}\left(t,t\right)-i\lambda\int_{0}^{\beta\hbar}d\tau\,\mathbb{G}^{\rceil}\left(t,\tau\right)\mathbf{V}_{M}\left(\tau\right)\mathbb{G}^{\lceil}\left(\tau,t\right)+\lambda\int_{t_{0}}^{t_{1}}dt^{\prime}\,\left[\mathbb{G}^{r}\left(t,t^{\prime}\right)\mathbf{V}_{-}\left(t^{\prime}\right)\mathbb{G}^{<}\left(t^{\prime},t\right)\right.\label{eq:G--_lesser}
\end{equation}
\[
\left.+\mathbb{G}^{<}\left(t,t^{\prime}\right)\mathbf{V}_{-}\left(t^{\prime}\right)\mathbb{G}^{a}\left(t^{\prime},t\right)+\mathbb{G}^{>}\left(t,t^{\prime}\right)\left(\mathbf{V}_{+}\left(t^{\prime}\right)-\mathbf{V}_{-}\left(t^{\prime}\right)\right)\mathbb{G}^{<}\left(t^{\prime},t\right)\right]
\]
and on the vertical track,
\[
\mathbf{G}_{MM}^{\lambda<}\left(\tau,\tau\right)\simeq\mathbb{G}_{MM}^{<}\left(\tau,\tau\right)-i\lambda\int_{0}^{\beta\hbar}d\tau^{\prime}\,\left[\mathbb{G}_{MM}^{r}\left(\tau,\tau^{\prime}\right)\mathbf{V}_{M}\left(\tau^{\prime}\right)\mathbb{G}_{MM}^{<}\left(\tau^{\prime},\tau\right)\right.
\]
\[
\left.+\mathbb{G}_{MM}^{<}\left(\tau,\tau^{\prime}\right)\mathbf{V}_{M}\left(\tau^{\prime}\right)\mathbb{G}_{MM}^{a}\left(\tau^{\prime},\tau\right)+\mathbb{G}_{MM}^{<}\left(\tau,\tau^{\prime}\right)\mathbf{V}_{M}\left(\tau^{\prime}\right)\mathbb{G}_{MM}^{>}\left(\tau^{\prime},\tau\right)\right]
\]
\begin{equation}
+\lambda\int_{t_{0}}^{t_{1}}dt\,\mathbb{G}^{\lceil}\left(\tau,t\right)\left(\mathbf{V}_{+}\left(t\right)-\mathbf{V}_{-}\left(t\right)\right)\mathbb{G}^{\rceil}\left(t,\tau\right)\label{eq:Gmm_lesser}
\end{equation}
Here, the dependence of the GF components on the coupling strength
is shown explicitly. As was already indicated, the unperturbed (electron-only,
i.e. without the coupling) GF $\mathbb{G}$ is defined for the Hamiltonian
which is the same on both horizontal tracks. Therefore, there is no
need to indicate the particular track anymore, and hence the $+/-$
subscripts have been omitted. Also, the right and left functions like
$\mathbb{G}_{+M}$ or $\mathbb{G}_{M-}$, which have one time on the
vertical and one on the horizontal tracks, do not require indicating
explicitly which horizontal track is used, and hence can simply be
denoted as $\mathbb{G}^{\rceil}$ and $\mathbb{G}^{\lceil}$, respectively. 

The first ($\lambda$ independent) term in the above expansions for
the three components of the GF gives rise to the first order approximation
to the effective action, Eq. (\ref{eq:eff-action-via-ful-GF}): 
\begin{equation}
\Delta S_{eff}^{(1)}\left[u(t),u^{\prime}(t),\overline{u}(\tau)\right]=i\hbar\,\text{tr}_{C}\left\{ \int_{t_{0}}^{t_{1}}dt\left[\mathbf{V}_{+}(t)-\mathbf{V}_{-}(t)\right]\mathbb{G}^{<}\left(t,t\right)-i\int_{0}^{\beta\hbar}d\tau\,\mathbf{V}_{M}\left(\tau\right)\mathbb{G}_{MM}^{<}\left(\tau,\tau\right)\right\} \label{eq:eff-action-1st-order-1}
\end{equation}
Using Eqs. (\ref{eq:V_mn-init}) and (\ref{eq:V_mn-t_gt_t0}) to express
the electron-nuclear matrix elements on different parts of $\gamma$
via atomic displacements, we obtain:
\[
\Delta S_{eff}^{(1)}\left[u(t),u^{\prime}(t),\overline{u}(\tau)\right]=\int_{t_{0}}^{t_{1}}dt\,\left[\sum_{A}i\hbar Y_{A}^{<}\left(t\right)\left(u_{A}(t)-u_{A}^{\prime}(t)\right)+\frac{i\hbar}{2}\sum_{AB}Y_{AB}^{<}(t)\left(u_{A}(t)u_{B}(t)-u_{A}^{\prime}(t)u_{B}^{\prime}(t)\right)\right]
\]
\begin{equation}
+\int_{0}^{\beta\hbar}d\tau\,\left[\sum_{A}\hbar\overline{Y}_{A}^{<}\left(\tau\right)\overline{u}_{A}\left(\tau\right)+\frac{\hbar}{2}\sum_{AB}\overline{Y}_{AB}^{<}\left(\tau\right)\overline{u}_{A}\left(\tau\right)\overline{u}_{B}\left(\tau\right)\right]\label{eq:action-1st-order-final}
\end{equation}
where 
\begin{equation}
Y_{A}^{<}\left(t\right)=\text{tr}_{C}\left[\mathbb{G}^{<}\left(t,t\right)\mathbf{V}^{A}\left(t\right)\right]\label{eq:Y-1st-order-A-t}
\end{equation}
\begin{equation}
Y_{AB}^{<}\left(t\right)=\text{tr}_{C}\left[\mathbb{G}^{<}\left(t,t\right)\mathbf{V}^{AB}\left(t\right)\right]\label{eq:Y-1st-order-AB-t}
\end{equation}
\begin{equation}
\overline{Y}_{A}^{<}\left(\tau\right)=\text{tr}_{C}\left[\mathbb{G}_{MM}^{<}\left(\tau,\tau\right)\mathbf{V}_{0}^{A}\right]\label{eq:Y-1st-order-A-tau}
\end{equation}
\begin{equation}
\overline{Y}_{AB}^{<}\left(\tau\right)=\text{tr}_{C}\left[\mathbb{G}_{MM}^{<}\left(\tau,\tau\right)\mathbf{V}_{0}^{AB}\right]\label{eq:Y-1st-order-AB-tau}
\end{equation}
Hence, the first order effective action contains both linear and quadratic
terms with respect to atomic displacements. Above, scalar functions
of real and imaginary times, Eqs. (\ref{eq:Y-1st-order-A-t})-(\ref{eq:Y-1st-order-AB-tau}),
have been defined.

Consider now the second order term which is obtained from the other
terms in Eqs. (\ref{eq:G++_lesser}), (\ref{eq:G--_lesser}) and (\ref{eq:Gmm_lesser}),
that are proportional to $\lambda$. Note that as the arguments of
the functions in the terms appearing in convolutions are implicit,
there is no need anymore to show them explicitly. We obtain after
some algebra:
\[
\Delta S_{eff}^{(2)}\left[u(t),u^{\prime}(t),\overline{u}(\tau)\right]=-\frac{i\hbar}{2}\int_{0}^{\beta\hbar}d\tau\int_{0}^{\beta\hbar}d\tau^{\prime}\,\text{tr}_{C}\left\{ \mathbf{V}_{M}\mathbb{G}_{MM}^{<}\mathbf{V}_{M}\left(\mathbb{G}_{MM}^{r}+\mathbb{G}_{MM}^{a}+\mathbb{G}_{MM}^{>}\right)\right\} 
\]
\[
+\frac{i\hbar}{2}\int_{t_{0}}^{t_{1}}dt\int_{t_{0}}^{t_{1}}dt^{\prime}\,\text{tr}_{C}\left[\left(\mathbb{G}^{r}+\mathbb{G}^{a}\right)\left(\mathbf{V}_{+}\mathbb{G}^{<}\mathbf{V}_{+}-\mathbf{V}_{-}\mathbb{G}^{<}\mathbf{V}_{-}\right)+\left(\mathbf{V}_{+}-\mathbf{V}_{-}\right)\left(\mathbb{G}^{>}\mathbf{V}_{+}\mathbb{G}^{<}-\mathbb{G}^{<}\mathbf{V}_{-}\mathbb{G}^{>}\right)\right]
\]
\begin{equation}
+\hbar\int_{0}^{\beta\hbar}d\tau\int_{t_{0}}^{t_{1}}dt\,\text{tr}_{C}\left[\left(\mathbf{V}_{+}-\mathbf{V}_{-}\right)\mathbb{G}^{\rceil}\mathbf{V}_{M}\mathbb{G}^{\lceil}\right]\label{eq:action-2nd-order-1}
\end{equation}

The transformations that follow are simple but rather cumbersome.
We first introduce new functions on the horizontal tracks,
\begin{equation}
v_{A}(t)=u_{A}(t)-u_{A}^{\prime}(t)\;\text{and}\;r_{A}(t)=\frac{1}{2}\left[u_{A}(t)+u_{A}^{\prime}(t)\right]\label{eq:new_displ_v+r}
\end{equation}
in terms of which the first and the second order contributions to
the action read:
\[
\Delta S_{eff}\left[u(t),u^{\prime}(t),\overline{u}(\tau)\right]=i\hbar\int_{t_{0}}^{t_{1}}dt\,\sum_{A}Y_{A}^{<}\left(t\right)v_{A}(t)+\hbar\int_{0}^{\beta\hbar}d\tau\,\sum_{A}\overline{Y}_{A}^{<}\left(\tau\right)\overline{u}_{A}\left(\tau\right)
\]
 
\[
+\frac{i\hbar}{2}\int_{t_{0}}^{t_{1}}dt\int_{t_{0}}^{t_{1}}dt^{\prime}\,\sum_{AB}\left\{ L_{AB}\left(t,t^{\prime}\right)v_{A}\left(t\right)v_{B}\left(t^{\prime}\right)+K_{AB}\left(t,t^{\prime}\right)r_{A}\left(t\right)v_{B}\left(t^{\prime}\right)\right\} 
\]
\begin{equation}
+\hbar\int_{0}^{\beta\hbar}d\tau\int_{t_{0}}^{t_{1}}dt\,\sum_{AB}Y_{AB}^{\rceil\lceil}\left(\tau,t\right)\overline{u}_{A}(\tau)v_{B}(t)-\frac{i\hbar}{2}\int_{0}^{\beta\hbar}d\tau\int_{0}^{\beta\hbar}d\tau^{\prime}\,\sum_{AB}\overline{L}_{AB}\left(\tau,\tau^{\prime}\right)\overline{u}_{A}\left(\tau\right)\overline{u}_{B}\left(\tau^{\prime}\right)\label{eq:action-2nd-order-2}
\end{equation}
where the following scalar functions were introduced:
\begin{equation}
L_{AB}\left(t,t^{\prime}\right)=\frac{1}{2}\left[Y_{BA}^{><}\left(t,t^{\prime}\right)+Y_{BA}^{<>}\left(t,t^{\prime}\right)\right]\equiv L_{BA}\left(t^{\prime},t\right)\label{eq:L_AB-coeff}
\end{equation}
\begin{equation}
K_{AB}\left(t,t^{\prime}\right)=2\delta\left(t-t^{\prime}\right)Y_{AB}^{<}(t)-2\theta_{t^{\prime}t}\left[Y_{BA}^{><}\left(t,t^{\prime}\right)-Y_{BA}^{<>}\left(t,t^{\prime}\right)\right]\label{eq:K_AB-coeff}
\end{equation}
\begin{equation}
\overline{L}_{AB}\left(\tau,\tau^{\prime}\right)=i\delta\left(\tau-\tau^{\prime}\right)\overline{Y}_{AB}^{<}\left(\tau\right)+\theta_{\tau\tau^{\prime}}\overline{Y}_{BA}^{<>}\left(\tau,\tau^{\prime}\right)+\theta_{\tau^{\prime}\tau}\overline{Y}_{BA}^{><}\left(\tau,\tau^{\prime}\right)\equiv\overline{L}_{BA}\left(\tau^{\prime},\tau\right)\label{eq:D_AB-coeff}
\end{equation}
where $\theta_{tt^{\prime}}$ ($\theta_{\tau\tau^{\prime}}$) is the
Heaviside function on the upper (vertical) track of the contour. Note
that the objects $L_{AB}\left(t,t^{\prime}\right)$ and $\overline{L}_{AB}\left(\tau,\tau^{\prime}\right)$
form symmetric matrices. Several $Y$ and $\overline{Y}$ double time
functions have also been defined:
\begin{equation}
Y_{AB}^{\alpha\beta}\left(t,t^{\prime}\right)=\text{tr}_{C}\left\{ \mathbb{G}^{\alpha}\left(t,t^{\prime}\right)\mathbf{V}^{A}\left(t^{\prime}\right)\mathbb{G}^{\beta}\left(t^{\prime},t\right)\mathbf{V}^{B}\left(t\right)\right\} \label{eq:Y_t,t'}
\end{equation}
\begin{equation}
\overline{Y}_{AB}^{\alpha\beta}\left(\tau,\tau^{\prime}\right)=\text{tr}_{C}\left\{ \mathbb{G}_{MM}^{\alpha}\left(\tau,\tau^{\prime}\right)\mathbf{V}_{0}^{A}\mathbb{G}_{MM}^{\beta}\left(\tau^{\prime},\tau\right)\mathbf{V}_{0}^{B}\right\} \label{eq:Y_bar-tau,tau'}
\end{equation}
\begin{equation}
Y_{AB}^{\rceil\lceil}\left(\tau,t\right)=\text{tr}_{C}\left\{ \mathbb{G}^{\rceil}\left(t,\tau\right)\mathbf{V}_{0}^{A}\mathbb{G}^{\lceil}\left(\tau,t\right)\mathbf{V}^{B}\left(t\right)\right\} \label{eq:Y-t-tau'}
\end{equation}
where $\alpha,\beta$ indicate various components $<$,$>$ of the
GFs contained in the trace. In simplifying the above expression use
have been made of the identities relating the retarded and advanced
GFs with the greater and lesser ones, $\mathbb{G}^{r}\left(z,z^{\prime}\right)=\theta_{zz^{\prime}}\left(\mathbb{G}^{>}\left(z,z^{\prime}\right)-\mathbb{G}^{<}\left(z,z^{\prime}\right)\right)$
and $\mathbb{G}^{a}\left(z,z^{\prime}\right)=-\theta_{z^{\prime}z}\left(\mathbb{G}^{>}\left(z,z^{\prime}\right)-\mathbb{G}^{<}\left(z,z^{\prime}\right)\right)$
which are valid when both arguments belong either to the horizontal
or vertical tracks of $\gamma$.

Concluding, we have got an expression for the effective action which
is found to be a second order form with respect to the functions $v_{A}(t)$,
$r_{A}(t)$ and $\overline{u}_{A}\left(\tau\right)$. When inserting
the effective action into the matrix element of the reduced density
matrix, Eq. (\ref{eq:final-path-int-expr-for-DM}), linear terms in
displacements can easily be incorporated into the existing actions
as they contain single time integrals; however, this is not the case
for quadratic in displacement terms as they enter via double time
integrals. Therefore, a procedure is required to linearilize them. 

\subsection{Real-imaginary time Hubbard-Stratanovich transformation\label{subsec:Hubbard-Stratanovich}}

Consider two sets of functions, $\left\{ k_{i}^{A}(t)\right\} $ and
$\left\{ \overline{k}_{j}^{A}(\tau)\right\} $, defined for real and
imaginary times, respectively, and the corresponding complex Gaussian
noises $\left\{ z_{i}^{A}(t)\right\} $ and $\left\{ \overline{z}_{j}^{A}(\tau)\right\} $.
A different number of functions may be in each set. Then, the following
identity can be established \cite{Gerard-PRB-2017} (see also \cite{Stockburger-PRL-2002}
where the same identity was written via contour integrals):
\[
\left\langle \exp\left\{ i\sum_{A}\left[\int_{t_{0}}^{t_{1}}dt\,\sum_{i}z_{i}^{A}(t)k_{i}^{A}(t)+\int_{0}^{\beta\hbar}d\tau\,\sum_{j}\overline{z}_{j}^{A}\left(\tau\right)\overline{k}_{j}^{A}\left(\tau\right)\right]\right\} \right\rangle _{\left\{ z,\overline{z}\right\} }
\]
\[
=\exp\left\{ -\frac{1}{2}\sum_{AB}\left[\int_{t_{0}}^{t_{1}}dt\int_{t_{0}}^{t_{1}}dt^{\prime}\,\sum_{ii^{\prime}}k_{i}^{A}(t)k_{i^{\prime}}^{B}\left(t^{\prime}\right)\varPi_{ii^{\prime}}^{AB}\left(t,t^{\prime}\right)+\int_{0}^{\beta\hbar}d\tau\int_{0}^{\beta\hbar}d\tau^{\prime}\,\sum_{jj^{\prime}}\overline{k}_{j}^{A}(\tau)\overline{k}_{j^{\prime}}^{B}\left(\tau^{\prime}\right)\overline{\varPi}_{jj^{\prime}}^{AB}\left(\tau,\tau^{\prime}\right)\right.\right.
\]
\begin{equation}
+\left.\left.2\int_{t_{0}}^{t_{1}}dt\int_{0}^{\beta\hbar}d\tau\,\sum_{j}\sum_{i}\overline{k}_{j}^{A}(\tau)k_{i}^{B}\left(t\right)\widetilde{\varPi}_{ji}^{AB}\left(\tau,t\right)\right]\right\} \label{eq:Hubbard-Stratanovich}
\end{equation}
where the indices $i$, $i^{\prime}$ were used for real time functions,
while $j$, $j^{\prime}$ for imaginary time functions. The introduced
double-time functions are actually correlation functions of the noises:
\begin{equation}
\varPi_{ii^{\prime}}^{AB}\left(t,t^{\prime}\right)=\left\langle z_{i}^{A}(t)z_{i^{\prime}}^{B}\left(t^{\prime}\right)\right\rangle _{\left\{ z,\overline{z}\right\} }\equiv\varPi_{i^{\prime}i}^{BA}\left(t^{\prime},t\right)\label{eq:corr-fun-tt'}
\end{equation}
\begin{equation}
\widetilde{\varPi}_{ji}^{AB}\left(\tau,t\right)=\left\langle \overline{z}_{j}^{A}(\tau)z_{i}^{B}\left(t\right)\right\rangle _{\left\{ z,\overline{z}\right\} }\label{eq:corr-fun-tau-t}
\end{equation}
\begin{equation}
\overline{\varPi}_{jj^{\prime}}^{AB}\left(\tau,\tau^{\prime}\right)=\left\langle \overline{z}_{j}^{A}(\tau)\overline{z}_{j^{\prime}}^{B}\left(\tau^{\prime}\right)\right\rangle _{\left\{ z,\overline{z}\right\} }\equiv\overline{\varPi}_{j^{\prime}j}^{BA}\left(\tau^{\prime},\tau\right)\label{eq:corr-fun-tau-tau'}
\end{equation}
Note that $\varPi_{ii^{\prime}}^{AB}\left(t,t^{\prime}\right)$ and
$\overline{\varPi}_{jj^{\prime}}^{AB}\left(\tau,\tau^{\prime}\right)$
form symmetric matrices. In the above equations each angle brackets
$\left\langle \ldots\right\rangle _{\left\{ z,\overline{z}\right\} }$
term corresponds to the stochastic averaging over the noises $\left\{ z,\overline{z}\right\} =\left\{ z_{i}^{A}(t),\overline{z}_{j}^{A}(\tau)\right\} $
with a Gaussian distribution function. These are actually path integrals
in their own right as the noises depend on time, see details in Ref.
\cite{Gerard-PRB-2017}. 

Similarly to the method developed in Ref. \cite{Gerard-PRB-2017},
for each $A$ we introduce four functions for the real time ($i=1,\ldots,4$)
and two functions for the imaginary one ($j=1,2$):
\begin{equation}
k^{A}(t)=\left(\begin{array}{c}
k_{1}^{A}(t)\\
k_{2}^{A}(t)\\
k_{3}^{A}(t)\\
k_{4}^{A}(t)
\end{array}\right)=\left(\begin{array}{c}
v_{A}(t)/\hbar\\
0\\
r_{A}(t)/\hbar\\
0
\end{array}\right),\quad\overline{k}^{A}(\tau)=\left(\begin{array}{c}
\overline{k}_{1}^{A}\left(\tau\right)\\
\overline{k}_{2}^{A}\left(\tau\right)
\end{array}\right)=\left(\begin{array}{c}
i\overline{u}_{A}\left(\tau\right)/\hbar\\
0
\end{array}\right)\label{eq:mapping-functions}
\end{equation}
which correspond to the noises as follows:
\begin{equation}
z^{A}(t)=\left(\begin{array}{c}
z_{1}^{A}(t)\\
z_{2}^{A}(t)\\
z_{3}^{A}(t)\\
z_{4}^{A}(t)
\end{array}\right)\equiv\left(\begin{array}{c}
\eta_{A}(t)\\
\eta_{A}^{*}(t)\\
\nu_{A}(t)\\
\nu_{A}^{*}(t)
\end{array}\right),\quad\overline{z}^{A}(\tau)=\left(\begin{array}{c}
\overline{z}_{1}^{A}\left(\tau\right)\\
\overline{z}_{2}^{A}\left(\tau\right)
\end{array}\right)\equiv\left(\begin{array}{c}
\overline{\mu}_{A}\left(\tau\right)\\
\overline{\mu}_{A}^{*}\left(\tau\right)
\end{array}\right)\label{eq:all-noises}
\end{equation}
The rational for choosing three pairs of complex conjugate noises
$\left(\eta_{A}(t),\eta_{A}^{*}(t)\right)$, $\left(\nu_{A}(t),\nu_{A}^{*}(t)\right)$
and $\left(\overline{\mu}_{A}(\tau),\overline{\mu}_{A}^{*}(\tau)\right)$
for each nuclear degree of freedom $A$ is as follows: (i) the noises
must be complex as their corresponding correlation functions are in
general complex (see below); (ii) we chose pairs of complex conjugate
noises to ensure that the Gaussian distribution associated with them
is real; our six complex noises are equivalent to choosing six \emph{real}
noises; (iii) three pairs of noises is a minimal possible set of noises
necessary to establish the mapping we need as for each $A$ there
are \emph{three} ``variables'' in the double integral (\ref{eq:action-2nd-order-2}):
$v_{A}(t)$, $r_{A}(t)$ and $\overline{u}_{A}(\tau)$. 

Then, the right hand side of Eq. (\ref{eq:Hubbard-Stratanovich})
can be exactly mapped to the second order effective action in Eq.
(\ref{eq:action-2nd-order-2}) if the following mapping conditions
are satisfied for the correlation functions of the noises:
\begin{equation}
\varPi_{11}^{AB}\left(t,t^{\prime}\right)\equiv\left\langle \eta_{A}(t)\eta_{B}\left(t^{\prime}\right)\right\rangle _{\left\{ z,\overline{z}\right\} }=\hbar^{2}L_{AB}\left(t,t^{\prime}\right)\label{eq:corr-function-eta-eta}
\end{equation}
\begin{equation}
\varPi_{31}^{AB}\left(t,t^{\prime}\right)\equiv\left\langle \nu_{A}(t)\eta_{B}\left(t^{\prime}\right)\right\rangle _{\left\{ z,\overline{z}\right\} }=\frac{\hbar^{2}}{2}K_{AB}\left(t,t^{\prime}\right)\label{eq:corr-function-nu-eta}
\end{equation}
\begin{equation}
\widetilde{\varPi}_{11}^{AB}\left(\tau,t\right)\equiv\left\langle \overline{\mu}_{A}(\tau)\eta_{B}\left(t\right)\right\rangle _{\left\{ z,\overline{z}\right\} }=-\hbar^{2}Y_{AB}^{\rceil\lceil}\left(\tau,t\right)\label{eq:corr-function-eta-mu}
\end{equation}
\begin{equation}
\overline{\varPi}_{11}^{AB}\left(\tau,\tau^{\prime}\right)\equiv\left\langle \overline{\mu}_{A}(\tau)\overline{\mu}_{B}\left(\tau^{\prime}\right)\right\rangle _{\left\{ z,\overline{z}\right\} }=\hbar^{2}\overline{L}_{AB}\left(\tau,\tau^{\prime}\right)\label{eq:corr-function-mu-mu}
\end{equation}
\begin{equation}
\varPi_{33}^{AB}\left(t,t^{\prime}\right)\equiv\left\langle \nu_{A}(t)\nu_{B}\left(t^{\prime}\right)\right\rangle _{\left\{ z,\overline{z}\right\} }=0,\quad\widetilde{\varPi}_{13}^{AB}\left(\tau,t\right)\equiv\left\langle \overline{\mu}_{A}(\tau)\nu_{B}\left(t\right)\right\rangle _{\left\{ z,\overline{z}\right\} }=0\label{eq:corr-functions-zero}
\end{equation}
Other correlation functions are not required. Note that the correlation
functions $\varPi_{11}^{AB}\left(t,t^{\prime}\right)$ and $\overline{\varPi}_{11}^{AB}\left(\tau,\tau^{\prime}\right)$
are automatically symmetric, as required, due to the symmetry of the
objects $L_{AB}\left(t,t^{\prime}\right)$ and $\overline{L}_{AB}\left(\tau,\tau^{\prime}\right)$,
respectively (see Eqs. (\ref{eq:L_AB-coeff}) and (\ref{eq:D_AB-coeff})).

We see that the noises satisfy certain correlations, which are related
to the electronic Green's functions. The correlation functions above
do not necessarily depend on the time difference; most likely they
depend on both times. It is seen that the correlation functions are
complex, see Eqs. (\ref{eq:L_AB-coeff})-(\ref{eq:Y-t-tau'}), which
justifies the choice we have made for the noises (\ref{eq:all-noises})
to be complex. 

\subsection{Nuclei only (reduced) density matrix\label{subsec:Nuclei-only-reduced-DM}}

Because of the established mapping, the contribution of the second
order effective action can be replaced with an average $\left\langle \cdots\right\rangle _{\left\{ z,\overline{z}\right\} }$
of a product of three fully independent exponential terms:
\[
\exp\left\{ \frac{i}{\hbar}\int_{t_{0}}^{t_{1}}dt\,\sum_{A}\left(\eta_{A}(t)+\frac{\nu_{A}(t)}{2}\right)u_{A}(t)\right\} \exp\left\{ -\frac{i}{\hbar}\int_{t_{0}}^{t_{1}}dt\,\sum_{A}\left(\eta_{A}(t)-\frac{\nu_{A}(t)}{2}\right)u_{A}^{\prime}(t)\right\} 
\]
\[
\times\exp\left\{ -\frac{1}{\hbar}\int_{0}^{\beta\hbar}d\tau\,\sum_{A}\overline{\mu}_{A}\left(\tau\right)\overline{u}_{A}(\tau)\right\} 
\]
Therefore, Eq. (\ref{eq:final-path-int-expr-for-DM}) can now be rewritten
as the average over the Gaussian noises,
\begin{equation}
\widehat{\rho}_{ions}\left(t_{1}\right)=\left\langle \widehat{\rho}_{ions}^{S}\left(t_{1}\right)\right\rangle _{\left\{ z,\overline{z}\right\} }\label{eq:Rho-via_average-Rho}
\end{equation}
where the superscript $S$ indicates that the density matrix operator
$\widehat{\rho}_{ions}^{S}$ corresponds to a particular manifestation
of the nosises, and 
\[
\left.\left\langle x_{1}\left|\widehat{\rho}_{ions}^{S}\left(t_{1}\right)\right|\right.x_{0}\right\rangle =\int dx_{2}dx_{3}\left\langle x_{1}\left.\left|\int\mathcal{D}x(t)\,e^{iS_{1}^{+}/\hbar}\right|x_{2}\right\rangle \right.
\]
\begin{equation}
\times\,\left\langle x_{2}\left.\left|\frac{Z_{2}^{0}}{Z_{0}}\int\mathcal{D}\overline{x}\left(\tau\right)\,e^{-\overline{S}_{1}/\hbar}\right|x_{3}\right\rangle \right.\,\left\langle \left.x_{3}\left|\int\mathcal{D}x^{\prime}(t)\,e^{-iS_{1}^{-}/\hbar}\right|x_{0}\right\rangle \right.\label{eq:final-path-int-expr-for-DM-1}
\end{equation}
is its corresponding matrix element. It contains the following real
and imaginary times actions acting on each track of $\gamma$, 
\begin{equation}
S_{1}^{+}=\int_{t_{0}}^{t_{1}}dt\,\left\{ L_{1}(t)+\sum_{A}\left[i\hbar Y_{A}^{<}\left(t\right)+\left(\eta_{A}(t)+\frac{\nu_{A}(t)}{2}\right)\right]u_{A}(t)\right\} \label{eq:Action_+}
\end{equation}
\begin{equation}
S_{1}^{-}=\int_{t_{0}}^{t_{1}}dt\,\left\{ L_{1}(t)+\sum_{A}\left[i\hbar Y_{A}^{<}\left(t\right)+\left(\eta_{A}(t)-\frac{\nu_{A}(t)}{2}\right)\right]u_{A}^{\prime}(t)\right\} \label{eq:Action_-}
\end{equation}
\begin{equation}
\overline{S}_{1}=\int_{0}^{\beta\hbar}d\tau\,\left\{ \overline{L}_{1}\left(\tau\right)+\sum_{A}\left[-i\hbar\overline{Y}_{A}^{<}\left(\tau\right)+\overline{\mu}_{A}\left(\tau\right)\right]\overline{u}_{A}\left(\tau\right)\right\} \label{eq:Action-tau}
\end{equation}
where $L_{1}(t)$ is the Lagrangian of an isolated nuclear subsystem,
$\overline{L}_{1}\left(\tau\right)$ being its corresponding Euclidean
counterpart; the electron-nuclear coupling enters here via the first
order and noise terms (expressions with the square brackets). The
actions above imply the following \emph{effective Hamiltonians} acting
on each track of the contour $\gamma$: 
\begin{equation}
\widehat{H}_{\pm}(t)=\mathcal{H}_{1}(t)-\sum_{A}\left[i\hbar Y_{A}^{<}\left(t\right)+\left(\eta_{A}(t)\pm\frac{\nu_{A}(t)}{2}\right)\right]u_{A}\label{eq:Hamiltonians-on-horisontal}
\end{equation}
\begin{equation}
\overline{H}\left(\tau\right)=\mathcal{H}_{1}^{0}+\sum_{A}\left[-i\hbar\overline{Y}_{A}^{<}\left(\tau\right)+\overline{\mu}_{A}\left(\tau\right)\right]u_{A}\label{eq:Hamiltoninan-on-vertical}
\end{equation}
where $u_{A}=x_{A}-\left\langle x_{A}\right\rangle _{t}$ for the
real time Hamiltonians $\widehat{H}_{\pm}(t)$, while $u_{A}=x_{A}-x_{A}^{0}$
for the imaginary time one, $\overline{H}\left(\tau\right)$.

In Eq. (\ref{eq:final-path-int-expr-for-DM-1}) the density matrix
is factorised and that enables one to write an\emph{ exact } expression
for the reduced (nuclei only) density matrix operator (see Appendix
A and Ref. \cite{Gerard-PRB-2017}), for the given manifestation of
the noises, as follows:
\begin{equation}
\widehat{\rho}_{ions}^{S}\left(t\right)=\widehat{U}_{+}\left(t,t_{0}\right)\widehat{\rho}_{0}^{S}\widehat{U}_{-}\left(t_{0},t\right)\label{eq:reduced-DM-operator}
\end{equation}
where $\widehat{\rho}_{0}^{S}\equiv\widehat{\rho}_{ions}^{S}\left(t_{0}\right)$
is the initial reduced density matrix (for the same noises), to be
discussed below, and two propagation operators have been introduced
on each part of the horizontal track 
\begin{equation}
\widehat{U}_{+}\left(t,t_{0}\right)=\widehat{\mathcal{T}}_{+}\exp\left\{ -\frac{i}{\hbar}\int_{t_{0}}^{t}dt^{\prime}\,\widehat{H}_{+}\left(t^{\prime}\right)\right\} \label{eq:U_upper}
\end{equation}
\begin{equation}
\widehat{U}_{-}\left(t_{0},t\right)=\widehat{\mathcal{T}}_{-}\exp\left\{ \frac{i}{\hbar}\int_{t_{0}}^{t}dt^{\prime}\,\widehat{H}_{-}\left(t^{\prime}\right)\right\} \label{eq:U_lower}
\end{equation}
which satisfy the following equations of motion:
\begin{equation}
i\hbar\partial_{t}\widehat{U}_{+}\left(t,t_{0}\right)=\widehat{H}_{+}(t)\widehat{U}_{+}\left(t,t_{0}\right)\label{eq:U_upper_EoM}
\end{equation}
\begin{equation}
i\hbar\partial_{t}\widehat{U}_{-}\left(t_{0},t\right)=-\widehat{U}_{-}\left(t_{0},t\right)\widehat{H}_{-}(t)\label{eq:U_lower_EoM}
\end{equation}
In the above equations $\widehat{\mathcal{T}}_{+}$ and $\widehat{\mathcal{T}}_{-}$
are the corresponding time-ordering operators on the upper (forward)
and lower (backward) tracks of the contour $\gamma$. Note that the
density matrix operator (\ref{eq:reduced-DM-operator}) is not normalised
to unity at any time $t$, $\text{Tr}_{N}\left(\widehat{\rho}_{ions}^{S}\left(t\right)\right)\neq1$,
although the final density matrix (\ref{eq:Rho-via_average-Rho})
is, $\text{Tr}_{N}\left(\widehat{\rho}_{ions}\left(t\right)\right)=1$
(the trace is understood here as being taken over the Hilbert subspace
of the nuclei subsystem).

\subsection{Initial preparation of the system}

In order to obtain the density matrix at real times, $\widehat{\rho}_{ions}^{S}\left(t\right)$
(for the given manifestation of the noises), one has to first determine
the corresponding initial density matrix $\widehat{\rho}_{0}^{S}$.
It can be found as the result of the propagation of an auxiliary density
matrix $\widehat{\rho}^{S}\left(\tau\right)$ in imaginary time,
\begin{equation}
\widehat{\rho}_{0}^{S}\equiv\left.\widehat{\rho}^{S}\left(\tau\right)\right|_{\tau=\beta\hbar},\quad\widehat{\rho}^{S}\left(\tau\right)=\frac{Z_{2}^{0}}{Z_{0}}\widehat{U}^{S}\left(\tau\right)=\xi\widehat{U}^{S}\left(\tau\right)\label{eq:initial-RDM-boundary-cond}
\end{equation}
where the imaginary time propagator 
\begin{equation}
\widehat{U}^{S}(\tau)=\overline{\mathcal{T}}\exp\left\{ -\frac{1}{\hbar}\int_{0}^{\tau}d\tau^{\prime}\,\overline{H}\left(\tau^{\prime}\right)\right\} \label{eq:imaginary-time-propagator}
\end{equation}
has been defined, $\widehat{U}^{S}\left(0\right)=\widehat{1}$, that
satisfies the following equation of motion in imaginary time:
\begin{equation}
-\hbar\partial_{\tau}\widehat{U}^{S}(\tau)=\overline{H}\left(\tau\right)\widehat{U}^{S}(\tau)\label{eq:Propagator-in-tau}
\end{equation}
with $\overline{\mathcal{T}}$ being the corresponding time-ordering
operators on the vertical (down) track of the contour $\gamma$. Of
course, the operator $\widehat{\rho}^{S}\left(\tau\right)$ is not
normalised to unity for any $\tau$.

At $t=t_{0}$ we have the boundary condition, 
\begin{equation}
\widehat{\rho}_{ions}^{S}\left(t_{0}\right)=\widehat{\rho}_{0}^{S}\equiv\left.\widehat{\rho}^{S}\left(\tau\right)\right|_{\tau=\beta\hbar}\label{eq:Boundary-Condition-for-Rho}
\end{equation}

Note that here $\widehat{\rho}_{ions}^{S}\left(t_{0}\right)$ (or
$\widehat{\rho}_{0}^{S}$) is defined up to (yet unknown) scaling
factor $\xi$, Eq. (\ref{eq:initial-RDM-boundary-cond}). This factor
can be obtained by noticing that the \emph{exact initial density matrix
}of the nuclear subsystem is obtained after averaging over the noises
$\left\{ \overline{\mu}_{A}\left(\tau\right)\right\} $, 
\begin{equation}
\widehat{\rho}_{0}\equiv\widehat{\rho}_{ions}\left(t_{0}\right)=\left\langle \widehat{\rho}_{0}^{S}\right\rangle _{\left\{ \overline{\mu}_{A}\right\} }\label{eq:Exact-initial-RDM}
\end{equation}
Note that correlations with the real time noises are irrelevant in
the case of the initial equilibration. Then, a subsequent normalisation
of the exact inital density matrix operator $\widehat{\rho}_{0}$
should fix the scaling factor $\xi$. 

Since the constant pre-factor $\xi=Z_{2}^{0}/Z_{0}$ does not depend
on the noises (it only depends on the initial Hamiltonian) and hence
is the same for each evolution of $\widehat{\rho}_{ions}^{S}\left(t\right)$,
i.e. for any particular manifestation of the noises, it can be determined
in practice by running a certain number of \emph{representative} imaginary
time evolutions as described below. 

From this point on we shall be using a matrix representation of the
density matrix operators and related quantities by employing an appropriate
basis set $\left\{ \chi_{i}(x)\right\} $ that depends on all nuclear
coordinates $x$. To obtain the numerical prefactor $\xi$ and the
initial density matrix, it is convenient to propagate numerically
the matrix $\mathbf{U}^{S}(\tau)=\left(U_{ij}^{S}\left(\tau\right)\right)$
associated with the auxiliary operator $\widehat{U}^{S}\left(\tau\right)$
(we shall remove the hat from operators when indicating their matrix
representation): 
\begin{equation}
-\hbar\partial_{\tau}U_{ij}^{S}(\tau)=\sum_{k}\overline{H}_{ik}\left(\tau\right)U_{kj}^{S}(\tau)\label{eq:DM-matrix-EoM}
\end{equation}
where 
\begin{equation}
\overline{H}_{ik}\left(\tau\right)=\left\langle \chi_{i}\right|\mathcal{H}_{1}^{0}\left|\chi_{k}\right\rangle +\sum_{A}\left[-i\hbar\overline{Y}_{A}^{<}\left(\tau\right)+\overline{\mu}_{A}\left(\tau\right)\right]\left\langle \chi_{i}\right|u_{A}\left|\chi_{k}\right\rangle \label{eq:DM-eq-H1-matrix}
\end{equation}
is the matrix element of the Hamiltonian (\ref{eq:Hamiltoninan-on-vertical}).
One must use the unit matrix as the initial condition in solving these
equations, $\mathbf{U}^{S}(\tau=0)=\mathbf{1}=\left(\delta_{ij}\right)$.
Then the normalisation factor is obtained via $\xi=1/\left\langle \xi_{0}^{S}\right\rangle _{\left\{ \overline{\mu}_{A}\right\} }$,
where $\xi_{0}^{S}=\text{tr}\left(\mathbf{U}^{S}\left(\beta\hbar\right)\right)$
is the trace of the auxiliary matrix corresponding to the given run.
Note that this calculation, if required, enables one also to determine
the matrix corresponding to the exact initial density matrix $\rho^{0}=\left(\rho_{ij}^{0}\right)$,
where $\rho_{ij}^{0}=\left\langle \chi_{i}\right|\widehat{\rho}_{0}\left|\chi_{j}\right\rangle $,
as $\rho_{ij}^{0}=\xi\left\langle U_{ij}^{S}\right\rangle _{\left\{ \overline{\mu}_{A}\right\} }$;
it is now properly normalised, $\text{tr}\left(\rho^{0}\right)=1$. 

Having obtained the normalisation factor, the real time simulations
of the matrix $\rho_{ions}(t)=\left(\left\langle \chi_{i}\right|\widehat{\rho}_{ions}(t)\left|\chi_{j}\right\rangle \right)$
can be initiated. We first run the imaginary time evolution to $\tau=\beta\hbar$
starting from the auxiliary density matrix with elements $\rho^{S}(0)=\left(\rho_{ij}^{S}(0)\right)=\left(\xi\delta_{_{ij}}\right)$.
Using the obtained matrix $\left(\rho_{ij}^{S}(\beta\hbar)\right)$
as the initial condition for $\rho_{ions}^{S}\left(t_{0}\right)$,
one proceeds with the real time run. This procedure ensures that,
at any time $t\geq t_{0}$, the density matrix $\left\langle \rho_{ions}^{S}(t)\right\rangle _{\left\{ z,\overline{z}\right\} }$,
sampled over all noises, will be properly normalised. Sampling over
all the real time runs, the total reduced density matrix, $\rho_{ions}(t)$,
at any time $t\geq t_{0}$ is obtained. 

Alternatively, one may run many imaginary + real time simulations
from $\rho^{S}(0)=\left(\rho_{ij}^{S}(0)\right)=\left(\delta_{_{ij}}\right)$,
and then determine the normalisation factor $\xi$ by sampling the
density matrix at some particular time $t\geq t_{0}$ and then normalising.
Then the actual reduced density matrix $\rho_{ions}\left(t\right)$
is obtained by scaling the calculated density at all times by $\xi$.

In any case, the expectation value $\left\langle \widehat{O}\right\rangle _{t}$
of any nuclear-only operator $\widehat{O}$ is calculated by taking
the trace of the product of the matices $\rho_{ions}\left(t\right)$
(after normalisation) and $\mathbf{O}=\left(O_{ij}\right)$, 
\[
\left\langle \widehat{O}\right\rangle _{t}=\sum_{ij}\left(\rho_{ions}\left(t\right)\right)_{ij}O_{ji}=\text{tr}\left(\rho_{ions}\left(t\right)\mathbf{O}\right)
\]

Equations (\ref{eq:reduced-DM-operator}) and (\ref{eq:initial-RDM-boundary-cond})
present the central result of this paper. They perform \emph{an exact
transition} from the path integrals of the reduced density matrix
of nuclei to its operator form. The path integrals were used as an
intermediate device to introduce the stochastic fields, and, by means
of the real-imaginary time Hubbard-Stratanovich transformation, to
factorize the influence functional and hence to make the reverse transformation
to the operator representation possible. 

\subsection{Equations of motion for nuclei\label{subsec:Equations-of-motion}}

Direct propagation of the observables in real time is also possible.
This may be numerically more preferable since the density matrix scales
quadratically with the number $n$ of the nuclear degrees of freedom,
while the number of observables will scale linearly with $n$. We
assume in what follows that the normalisation prefactor $\xi$ is
known, and hence the initial density matrix $\rho_{ions}^{S}\left(t_{0}\right)$,
obtained by propagating in imaginary time the auxiliary matrix $\rho^{S}\left(\tau\right)$,
is also known. 

Having obtained an explicit expression for the operator of the density
matrix of the nuclei, Eq. (\ref{eq:reduced-DM-operator}), we can
differentiate it with respect to time. Using the equations of motion
for the propagation operators, Eqs. (\ref{eq:U_upper_EoM}) and (\ref{eq:U_lower_EoM}),
one can easily obtain an equation of motion for the reduced density
matrix:
\[
i\hbar\partial_{t}\widehat{\rho}_{ions}^{S}(t)=\widehat{H}_{+}(t)\widehat{\rho}_{ions}^{S}(t)-\widehat{\rho}_{ions}^{S}(t)\widehat{H}_{-}(t)
\]
\begin{equation}
=\left[\mathcal{H}_{1}\left(t\right),\widehat{\rho}_{ions}^{S}(t)\right]_{-}-\sum_{A}\left\{ \left(i\hbar Y_{A}^{<}\left(t\right)+\eta_{A}(t)\right)\left[x_{A},\widehat{\rho}_{ions}^{S}(t)\right]_{-}+\frac{\nu_{A}(t)}{2}\left[u_{A}(t),\widehat{\rho}_{ions}^{S}(t)\right]_{+}\right\} \label{eq:EoM-for-RDM}
\end{equation}
where $u_{A}(t)=x_{A}-\left\langle x_{A}\right\rangle _{t}^{S}$ is
the displacement operator. The equation of motion contains both commutators
and anticommutators, indicated with the minus and plus subscripts,
respectively. As expected \cite{Stockburger-PRL-2002,Gerard-PRB-2017},
the dynamical evolution of the density matrix of an open system (the
nuclei) contains the anitcommutator and hence is not Hamiltonian.

Correspondingly, an equation of motion for the expectation value of
an arbitrary\emph{ nuclear-only} operator $\widehat{O}$ (for a particular
realization of the noises, indicated again by the superscript $S$)
reads:
\begin{equation}
i\hbar\partial_{t}\left\langle \widehat{O}\right\rangle _{t}^{S}=\left\langle \left[\widehat{O},\mathcal{H}_{1}(t)\right]_{-}\right\rangle _{t}^{S}-\sum_{A}\left\{ \left(i\hbar Y_{A}^{<}\left(t\right)+\eta_{A}(t)\right)\left\langle \left[\widehat{O},x_{A}\right]_{-}\right\rangle _{t}^{S}+\frac{\nu_{A}(t)}{2}\left\langle \left[\widehat{O},u_{A}(t)\right]_{+}\right\rangle _{t}^{S}\right\} \label{eq:EoM-for-operator-O}
\end{equation}
where $\left\langle \widehat{O}\right\rangle _{t}^{S}=\text{Tr}_{N}\left(\widehat{\rho}_{ions}^{S}(t)\widehat{O}\right)$.
As one can see, this equation contains only real time noises; however,
it is to be emphasized that the real time noises are correlated with
the imaginary time noises used to generate the initial auxiliary density
matrix. In general, if transient effects (immediately after initial
equilibration) are of interest, these correlations must be taken into
account in calculating the real time evolution. 

Applying Eq. (\ref{eq:EoM-for-operator-O}) to the unity operator,
$\widehat{O}=\widehat{1}$, one obtains an equation for the time evolution
of the trace $\xi^{S}(t)=\text{Tr}_{N}\left(\widehat{\rho}_{ions}^{S}(t)\right)$
of the density matrix for the given stochastic run:
\[
\partial_{t}\xi^{S}(t)=-\frac{1}{i\hbar}\sum_{A}\nu_{A}(t)\left\langle u_{A}(t)\right\rangle _{t}^{S}
\]
Since by definition $u_{A}(t)=x_{A}-\left\langle x_{A}\right\rangle _{t}^{S}$,
then $\left\langle u_{A}(t)\right\rangle _{t}^{S}=0$. This means
that the trace of the density matrix remains constant during each
stochastic run in real time, being equal to its initial value, $\xi^{S}(t)\equiv\xi_{0}^{S}$,
at time $t_{0}$. Of course, these values will be different for different
runs.

We shall now apply the above result to derive the time evolution of
the nuclear positions, when $\widehat{O}\rightarrow x_{A}$. In this
case we obtain:
\begin{equation}
\partial_{t}\left\langle x_{A}\right\rangle _{t}^{S}=\frac{1}{m_{A}}\left\langle p_{A}\right\rangle _{t}^{S}-\frac{1}{i\hbar}\sum_{B}\nu_{B}\left(t\right)\left[\left\langle x_{A}x_{B}\right\rangle _{t}^{S}-\left\langle x_{A}\right\rangle _{t}^{S}\left\langle x_{B}\right\rangle _{t}^{S}\right]\label{eq:EoM-for-x_A}
\end{equation}
where $p_{A}=-i\hbar\frac{\partial}{\partial x_{A}}$ is the corresponding
momentum operator. Note that an expression within the square brackets
is in fact a fluctuation $\left\langle u_{A}(t)u_{B}(t)\right\rangle _{t}^{S}$. 

Putting $\widehat{O}\rightarrow p_{A}$ in Eq. (\ref{eq:EoM-for-operator-O}),
we obtain a complimentary equation of motion for the momentum:
\begin{equation}
\partial_{t}\left\langle p_{A}\right\rangle _{t}^{S}=-\left\langle \frac{\partial U_{1}}{\partial x_{A}}\right\rangle _{t}^{S}+\left(i\hbar Y_{A}^{<}\left(t\right)+\eta_{A}(t)+\frac{\nu_{A}(t)}{2}\right)\xi_{0}^{S}-\frac{1}{i\hbar}\sum_{B}\nu_{B}\left(t\right)\left[\left\langle x_{B}p_{A}\right\rangle _{t}^{S}-\left\langle x_{B}\right\rangle _{t}^{S}\left\langle p_{A}\right\rangle _{t}^{S}\right]\label{eq:EoM-for-p_A}
\end{equation}
where $U_{1}$ is the potential energy of nuclei due to the nuclei-nuclei
interaction as well as due to a possible external field acting directly
on nuclei. The last term in the square brackets contains another fluctuation,
$\left\langle u_{B}(t)p_{A}\right\rangle _{t}^{S}$. 

To proceed, we have to find an appropriate expression for the derivative
of the potential $U_{1}$. In principle, one can try to write an equation
of motion for the operator $\partial U_{1}/\partial x_{A}$. It depends
only on the atomic positions and hence only its commutator with momenta
in the first term of Eq. (\ref{eq:EoM-for-operator-O}) need to be
considered, as well as the last - anticommutator - term in the same
equation. The equation obtained in this way would also contain noises
in its right hand side:
\[
\partial_{t}\left\langle \frac{\partial U_{1}}{\partial x_{A}}\right\rangle _{t}^{S}=\sum_{B}\frac{1}{m_{B}}\left\langle \frac{\partial^{2}U_{1}}{\partial x_{A}\partial x_{B}}\right\rangle _{t}^{S}-\frac{1}{i\hbar}\sum_{B}\nu_{B}\left(t\right)\left[\left\langle \frac{\partial U_{1}}{\partial x_{A}}x_{B}\right\rangle _{t}^{S}-\left\langle \frac{\partial U_{1}}{\partial x_{A}}\right\rangle _{t}^{S}\left\langle x_{B}\right\rangle _{t}^{S}\right]
\]
As a way of illustration, let us also work out this term within the
harmonization approximation \cite{LK-c_number-PRB-2017}. Indeed,
expanding the potential energy $U_{1}$ in terms of nuclear displacements
up to the second order,
\[
U_{1}\simeq U_{1}^{0}+\sum_{A}\frac{\partial U_{1}\left(\left\langle x\right\rangle _{t}^{S}\right)}{\partial\left\langle x_{A}\right\rangle _{t}^{S}}u_{A}(t)+\frac{1}{2}\sum_{AB}\frac{\partial^{2}U_{1}\left(\left\langle x\right\rangle _{t}^{S}\right)}{\partial\left\langle x_{A}\right\rangle _{t}^{S}\partial\left\langle x_{B}\right\rangle _{t}^{S}}u_{A}(t)u_{B}(t)
\]
we obtain after differentiation and taking the average:
\begin{equation}
\left\langle \frac{\partial U_{1}}{\partial x_{A}}\right\rangle _{t}^{S}\simeq\frac{\partial U_{1}\left(\left\langle x\right\rangle _{t}^{S}\right)}{\partial\left\langle x_{A}\right\rangle _{t}^{S}}\label{eq:average-of-force}
\end{equation}
If we expand $U_{1}$ to the third order in the displacements, which
is beyond the harmonization approximation, then the next term would
contain fluctuations $\left\langle u_{B}(t)u_{C}(t)\right\rangle _{t}^{S}$.
One can see that within the harmonization approximation the average
(\ref{eq:average-of-force}) can easily be calculated directly from
the mean atomic positions; no need in this case to construct a specific
equation of motion for the average. 

The obtained equations are not self-contained since they require knowledge
of the time evolution of additional quantities such as $\left\langle x_{A}x_{B}\right\rangle _{t}^{S}$,
$\left\langle x_{A}p_{B}\right\rangle _{t}^{S}$, $\left\langle \frac{\partial^{2}U_{1}}{\partial x_{A}\partial x_{B}}\right\rangle _{t}^{S}$,
and $\left\langle \frac{\partial U_{1}}{\partial x_{A}}x_{B}\right\rangle _{t}^{S}$.
Writing the corresponding equations of motion for these quantities
results in the appearance of higher order fluctuations. In the end,
one obtains an infinite set of hierarchical equations of motion containing
higher order fluctuations and higher order derivatives of the potential
$U_{1}$. Correspondingly, in practice one has to terminate the hierarchy
at some point to obtain a finite set of equations. 

Note that solution of the above equations requires knowledge of the
initial (at $t=t_{0}$) mean values of all operators these equations
contain. These are easily obtained, for each stochastic run, at the
end of the imaginary time evolution via 
\begin{equation}
\left\langle \widehat{O}\right\rangle _{t=t_{0}}^{S}=\text{tr}\left(\rho_{ions}^{S}\left(t_{0}\right)\mathbf{O}\right)\label{eq:DM-matrix-expectation-value}
\end{equation}

\section{Discussion and conclusions}

In this paper we have considered a coupled system of nuclei and electrons.
Either all or some of the nuclei are allowed to move. The goal was
to obtain equations of motion for the nuclei taking full account of
their interaction with the electrons and the electronic relaxation.
Our method is based on a few ideas developed in Refs. \cite{Hedegard-PRB-1987,Brandbyge-PRB-2012}.
However, at variance with the mentioned work, in our method we do
not invoke the partition approximation as the nuclei and electrons
are considered fully coupled and thermalized at the initial time,
which is more physically sound. In addition to that, we demonstrated,
following the previous work \cite{Gerard-PRB-2017}, how one can get
an exact expression for the reduced density matrix from its path integral
representation. 

The derived equations of motion for the atomic positions have a form
of an infinite hierarchy of stochastic differential equations, containing
three types of noises ($\left\{ \eta_{A}(t)\right\} $, $\left\{ \nu_{A}(t)\right\} $
and $\left\{ \overline{\mu}_{A}(\tau)\right\} $) in real and imaginary
times for each nuclear degree of freedom $A$ that are considered
explicitly (allowed to move). The noises are correlated with each
other via various components of the electronic-only GF $\mathbb{G}$.
We do not specify how this GF is to be calculated, this depends on
the particular problem at hand. However, our analysis seems to suggest
that no matter what kind of a system is actually considered, would
it be a molecule under an electric pulse or a molecular junction under
a bias, the general form of the equations of motion for the nuclei
remains the same; it is \emph{universal}. The particular problem under
consideration imprints on the GF $\mathbb{G}$ and hence on the correlation
functions of the noises to be considered to generate them. 

The calculation proceeds in the following way: 
\begin{enumerate}
\item During an imaginary time evolution, Eqs. (\ref{eq:DM-matrix-EoM})
and (\ref{eq:DM-eq-H1-matrix}), the system is initially prepared
(thermalized) adopting a certain nuclear basis set $\left\{ \chi_{i}(x)\right\} $.
One starts from the unit auxiliary matrix at $\tau=0$, i.e. $\left(U_{ij}\left(0\right)\right)=\mathbf{1}$,
and then propagates it in time up to $\tau=\beta\hbar$. The trace
of the obtained auxiliary matrix, $\xi_{0}^{S}$, is stored. This
calculation requires generating only one set of noises, $\left\{ \overline{\mu}_{A}(\tau)\right\} $,
that are correlated via Eq. (\ref{eq:corr-function-mu-mu}), and hence
the knowledge of the electronic GF on the vertical track of the contour
is only needed, $\mathbb{G}_{MM}^{<}\left(\tau,\tau^{\prime}\right)$
and $\mathbb{G}_{MM}^{>}\left(\tau,\tau^{\prime}\right)$, see Eqs.
(\ref{eq:Y-1st-order-AB-tau}) and (\ref{eq:Y_bar-tau,tau'}), as
only they determine the function $\overline{L}_{AB}\left(\tau,\tau^{\prime}\right)$,
Eq. (\ref{eq:D_AB-coeff}), which enters the correlation of the noise.
The calculations are repeated the necessary number of times for different
noises, and the normalisation prefactor, $\xi=1/\left\langle \xi_{0}^{S}\right\rangle _{\left\{ \overline{\mu}_{A}\right\} }$,
is worked out. The sampling is stopped when $\xi$ is converged (does
not change upon addition of new stochastic runs). 
\item To propagate the nuclear system in real time, one has to decide upon
terminatioon at a certain order of the hierarchy of stochastic differential
equations considered in section \ref{subsec:Equations-of-motion}
by setting any higher order fluctuations and derivatives of the nuclear
potential energy to zero. In the case of the potential energy this
approximation simply corresponds to adopting a Taylor expansion of
$U_{1}$ in terms of the nuclear displacements $u_{A}=x_{A}-\left\langle x_{A}\right\rangle _{t}^{S}$
terminated at a certain order; the harmonization approximation where
all terms after the quadratic ones are set to zero being the simplest
approximation. The equations adopted contain a finite number of specific
expectation values $\left\langle \widehat{O}_{1}\right\rangle _{t}^{S}$,
$\left\langle \widehat{O}_{2}\right\rangle _{t}^{S}$, etc. to propagate
in time.
\item Then, using the adopted matrix representation, many time evolutions
are run. Each such evolution consists of an imaginary run followed
up by the real time one. 
\begin{enumerate}
\item Diring the imaginary time run, one starts with the auxiliary matrix
$\rho^{S}(\tau=0)=\left(\xi\delta_{ij}\right)$ and then propagates
it up to $\tau=\beta\hbar$; then the value of the trace $\xi_{0}^{S}$
is stored and the initial values of all expectation values $\left\langle \widehat{O}_{1}\right\rangle _{t_{0}}^{S}$,
$\left\langle \widehat{O}_{2}\right\rangle _{t_{0}}^{S}$, etc. that
are met in the equations of motion are also calculated using Eq. (\ref{eq:DM-matrix-expectation-value}).
\item Then one has to propagate numerically those expectation values in
real time employing a small time step $\Delta t$ using the derived
equations of motion for them from the terminated hierarchy. This calculation
requires generating two sets of noises, $\left\{ \eta_{A}(t)\right\} $
and $\left\{ \nu_{A}(t)\right\} $, which are to be correlated not
only between themselves, but also with the noises $\left\{ \overline{\mu}_{A}(\tau)\right\} $
generated for the initial preparation of the system (point 3a above),
for each particular run. The calculation of the correlation functions
(\ref{eq:corr-function-eta-eta}) - (\ref{eq:corr-function-eta-mu})
requires obtaining all other components of the electronic GF $\mathbb{G}^{<}$,
$\mathbb{G}^{>}$, $\mathbb{G}^{\lceil}$, and $\mathbb{G}^{\rceil}$
at progressive times, see Eqs. (\ref{eq:L_AB-coeff}), (\ref{eq:K_AB-coeff}),
(\ref{eq:Y_t,t'}) and (\ref{eq:Y-t-tau'}). The GFs depend on the
actual positions of the atoms, $\left\langle x_{A}\right\rangle _{t}^{S}$,
as these modify the electronic Hamiltonian. Therefore, the numerical
solution of the equations of motion requires recalculating the GFs,
and hence the correlation functions and the noises $\left\{ \eta_{A}(t)\right\} $
and $\left\{ \nu_{A}(t)\right\} $, at each consecutive time step
(or after a certain number of such steps). The recalculation of the
components of the GF ($\mathbb{G}^{<}$, $\mathbb{G}^{>}$, $\mathbb{G}^{\rceil}$
and $\mathbb{G}^{\lceil}$) can be done by solving their corresponding
equations of motion (Kadanoff-Baym equations) numerically using e.g.
the time-stepping technique; there is a significant experience in
this regard (see e.g. the book \cite{Stefanucci-Leeuwen}, p. 472).
\end{enumerate}
\item The calculation is repeated with different realizations of the noises,
and the final result is obtained by sampling over all such calculations,
i.e.
\[
\left\langle x_{A}\right\rangle _{t}=\left\langle \,\left\langle x_{A}\right\rangle _{t}^{S}\,\right\rangle _{\left\{ \eta_{A},\nu_{A},\overline{\mu}_{A}\right\} },\quad\left\langle p_{A}\right\rangle _{t}=\left\langle \,\left\langle p_{A}\right\rangle _{t}^{S}\,\right\rangle _{\left\{ \eta_{A},\nu_{A},\overline{\mu}_{A}\right\} }
\]
and so on.
\end{enumerate}
The described algorithm enables one not only to calculate the mean
atomic trajectory $\left\langle x_{A}\right\rangle _{t}$ of atoms
in the system, but also various their fluctuations from the mean trajectory.
This naturally corresponds to the fact that nuclei in our method are
considered fully quantum-mechanically.

Basically, only two approximations have been made in our theory: (i)
the part of the Hamiltonian responsible for the interaction between
electrons and nuclei (in the full-electron picture that would simply
be the corresponding Coulomb interaction term) was considered up to
the second order with respect to the displacements of the nuclei from
either their instantaneous positions given by the mean atomic trajectory
(real time evolution) or from their equilibrium positions (initial
preparation corresponding to the imaginary time evolution, Eqs. (\ref{eq:initial-RDM-boundary-cond})
and (\ref{eq:Propagator-in-tau})); (ii) the electronic GF was expanded
only up to the first order with respect to such displacements, Eq.
(\ref{eq:GF-to-the-1st-order}). Both approximations are consistent
with each other and correspond to the harmonization picture. No other
approximations have been made. 

Note that the harmonization approximation applied here does not assume
small electron-nuclear interaction; instead, it adopts a view that
only small fluctuations of atomic positions from their either the
equilibrium positions (initial preparation of the system) or the mean
trajectory (during the real time evolution) are essential. This implies
an application of this method to not too large temperatures.

One may wonder how our method is related to Generalised Langevin Equation
(GLE) methods, see e,g. \cite{LK-c_number-PRB-2017} and references
therein, which can also be used to consider atomic trajectories of
nuclei treated as an open system (in that case the bath was considered
a set of harmonic oscillators, however). In GLE method equations of
motion for mean atomic trajectories have the form of stochastic differential
equations with a time integral term (friction), which corresponds
to a memory accumulated during preceding times. Formally, our equations
of motion have a very different form; for instance, there is no memory
time integrals at all in our equations. This observation, however,
may be too rushed: indeed, we have to deal with more than one equation:
there are in fact two sets. One consists of the equations for atomic
positions and momenta, and the other - of other equations for various
fluctuations and derivatives of the potential $U_{1}$. Therefore,
solving formally for all fluctuations from the second set (this solution
would have time integral terms) and substituting the obtained expressions
into the equations for the positions and momenta of the first set,
Eqs. (\ref{eq:EoM-for-x_A}) and (\ref{eq:EoM-for-p_A}), an integral
memory-like term would appear. However, at this stage these are just
rather general observations; detailed execution of the above program
is left for future work. 

The theoretical framework proposed in this paper is rather general.
Its application to a particular system is imprinted in the nuclear-only
potential energy $U_{1}$ and in the electronic GF. The form of the
equations of motion for atoms is, however, universal. This method
we hope will also contribute to the discussion \cite{DiVentra-book,Dundas-NatNano-2009,Bode-PRL-2011,Brandbyge-PRB-2012,vOppen-Belst-2012}
of current induced forces although, when the atomic nuclei are considered
as quantum, the notion of a classical ``atomic force'' is not strictly
meaningful (quantum nuclei were initially considered also in \cite{Brandbyge-PRB-2012}
although in the end the authors went back to classical nuclei when
writing a GLE for them).

The other point is that our method is fully numerical: to obtain the
atomic trajectory (and their fluctuations) one has to solve the equations
of motion discussed above on a computer. Therefore, this method may
not only be used for perform actual calculations on realistic systems;
it can serve as a benchmark for various approximate methods.

There are several important technical issues that require further
thorough investigation. The main fundamental point is related to the
level at which the hierarchy of equations of motion can be safely
terminated; surely this should depend on the problem at hand. The
other point is related to the time step to choose in order to integrate
the equations of motion numerically; we expect that this time step
may be longer than the electronic (femtosecond) range, but might be
shorter than the timescale of atomic vibrations. Finally, and most
importantly, an efficient implementation of the method must be developed.
The mentioned points require detailed investigations and are left
for future work.

Finally, we note that our method can only be used specifically for
obtaining atomic trajectories (or expectation values of any nuclear-related
operators). It cannot be used for calculating expectation values associated
with electronic operators such as the electronic density or current.
To obtain those along the mean trajectory of atoms as functions of
time, other methods need to be developed. These are presently being
developed in our laboratory.

We hope that this work will stimulate further research on atomic dynamics
of general non-equilibrium systems. 

\section*{Appendix A}

Consider a general propagation operator 
\[
\widehat{U}_{\gamma}\left(z,z^{\prime}\right)=\widehat{\mathcal{T}}_{\gamma}\exp\left\{ -\frac{i}{\hbar}\int_{z^{\prime}}^{z}\mathcal{H}\left(z_{1}\right)dz_{1}\right\} 
\]
between any two points $z,z^{\prime}\in\gamma$ on the contour shown
in Fig. \ref{fig:Konstantinov-Perel-contour}. In practice, to derive
Eq. (\ref{eq:red-DM-of-ions}), we shall only need to consider three
particular cases, where both times lie \emph{on the same }track: (i)
both $z=t$ and $z^{\prime}=t^{\prime}$ are on the upper track, $t>t^{\prime}$;
(ii) $z=t$ and $z^{\prime}=t^{\prime}$ are on the lower tack with
$z>z^{\prime}$ (meaning that $t<t^{\prime}$), and (iii) $z=t_{0}^{-}-i\tau$
and $z^{\prime}=t_{0}^{-}$ are on the vertical track with $z>z^{\prime}$
($\tau>0$). However, we proceed with the derivation in the general
case and will consider these particular cases at the end of the calculation.

The propagator satisfies the semi-group property, $\widehat{U}_{\gamma}\left(z,z^{\prime}\right)=\widehat{U}_{\gamma}\left(z,z_{1}\right)\widehat{U}_{\gamma}\left(z_{1},z^{\prime}\right)$
and also $\widehat{U}_{\gamma}\left(z,z\right)=1$. 

We are interested in writing the propagator in the coordinate representation
with respect to the nuclear positions $x_{0}$ and $x$ (from region
1). Let us split the part of the contour between $z^{\prime}$ and
$z$ by $n$ ``equidistant'' points with the distance between them
$\left|\epsilon\right|\sim1/n$. Note that the meaning of $\epsilon\equiv\Delta z$
depends on where $\Delta z$ is: it is equal to $\Delta t$ or $-\Delta t$
on the horizontal upper or lower tracks, respectively, and to $-i\Delta\tau$
on the vertical track. Then, writing the propagator as a product of
propagators over each interval, we obtain - by inserting the resolution
of identity $\int\left|x\right\rangle \left\langle x\right|dx=1$
in appropriate places - an expression:
\[
\left.\left\langle x_{f}\left|\widehat{U}_{\gamma}\left(z,z^{\prime}\right)\right|x_{0}\right.\right\rangle =\int dx_{n-1}\cdots dx_{1}\left.\left\langle x_{f}\left|\widehat{U}_{\gamma}\left(z,z_{n-1}\right)\right|x_{n-1}\right.\right\rangle \cdots\left.\left\langle x_{1\text{}}\left|\widehat{U}_{\gamma}\left(z_{1},z_{0}\right)\right|x_{0}\right.\right\rangle 
\]
Here $x_{j}$ is a vector of a particular instance of all nuclei degrees
of freedom associated with the time $z_{j}$ on the contour, with
$j=0,\ldots,n$. We set $z_{n}=z$ and $z_{0}=z^{\prime}$; also,
$x_{0}$ corresponds to $z^{\prime}$ and $x_{f}$ to $z$. 

The Hamiltonian in the propagator $\mathcal{H}=\left(K_{1}+K_{2}\right)+\left(V_{1}\left(x\right)+V_{12}\left(x\right)+V_{2}\right)$
consists of kinetic and potential energy terms for each region, as
well as of the interaction $V_{12}$ between them. For small $\epsilon$
every elementary propagator can be factorised, 
\[
\widehat{U}_{\gamma}\left(z_{j+1},z_{j}\right)\simeq e^{-i\epsilon\left(K_{1}+K_{2}\right)/\hbar}e^{-i\epsilon\left(V_{1}+V_{2}+V_{12}\right)/\hbar}
\]
leading to 
\[
\left.\left\langle x_{j+1}\left|\widehat{U}_{\gamma}\left(z_{j+1},z_{j}\right)\right|x_{j}\right.\right\rangle \simeq e^{-i\epsilon K_{2}/\hbar}e^{-i\epsilon\left(V_{1}\left(x_{j}\right)+V_{2}+V_{12}\left(x_{j}\right)\right)/\hbar}\left.\left\langle x_{j+1}\left|e^{-i\epsilon K_{1}/\hbar}\right|x_{j}\right.\right\rangle 
\]
The remaining matrix element is worked out in a usual way by inserting
twice the resolution of identity $\int\left|p\right\rangle \left\langle p\right|dp=1$
with respect to the vector $p$ of all nuclear momenta, and using
the fact that 
\[
\left.\left\langle x\right|p\right\rangle =\left(\prod_{A}\frac{1}{\sqrt{2\pi\hbar}}\right)e^{ipx/\hbar}
\]
and 
\[
\left\langle p\right|e^{-i\epsilon K_{1}/\hbar}\left|p^{\prime}\right\rangle =\delta\left(p-p^{\prime}\right)e^{-iK_{1}\left(p\right)\epsilon/\hbar}
\]
We obtain:
\[
\left\langle x_{j+1}\right|\widehat{U}_{\gamma}\left(z_{j+1},z_{j}\right)\left|x_{j}\right\rangle =\left[\prod_{A}\left(\frac{m_{A}}{2\pi\hbar i\epsilon}\right)^{1/2}\right]\exp\left\{ \frac{i\epsilon}{\hbar}\left[\sum_{A}\frac{m_{A}}{2}\left(\frac{x_{A,j+1}-x_{A,j}}{\epsilon}\right)^{2}-V_{1}\left(x_{j}\right)\right]\right\} 
\]
\[
\times\,\exp\left\{ -\frac{i\epsilon}{\hbar}K_{2}\right\} \exp\left\{ -\frac{i\epsilon}{\hbar}\left[V_{12}\left(x_{j}\right)+V_{2}\right]\right\} 
\]
where use has been made of the fact that the positions and momenta
of region 1 commute with those of region 2.

Correspondingly, the coordinate representation of the propagator is
a multiple integral of a product of the above type of terms calculated
for different positions $x_{j}$: 
\[
\left.\left\langle x_{f}\left|\widehat{U}_{\gamma}\left(z,z^{\prime}\right)\right|x_{0}\right.\right\rangle \simeq\int dx_{n-1}\cdots dx_{1}\left[\prod_{A}\left(\frac{m_{A}}{2\pi\hbar i\epsilon}\right)^{n/2}\right]\exp\left\{ \frac{i\epsilon}{\hbar}\sum_{j=0}^{n-1}\left[\sum_{A}\frac{m_{A}}{2}\left(\frac{x_{A,j+1}-x_{A,j}}{\epsilon}\right)^{2}-V_{1}\left(x_{j}\right)\right]\right\} 
\]
\begin{equation}
\times e^{-iK_{2}\epsilon/\hbar}e^{-i\left[V_{12}\left(x_{n-1}\right)+V_{2}\right]\epsilon/\hbar}\cdots e^{-iK_{2}\epsilon/\hbar}e^{-i\left[V_{12}\left(x_{j}\right)+V_{2}\right]\epsilon/\hbar}\cdots e^{-iK_{2}\epsilon/\hbar}e^{-i\left[V_{12}\left(x_{0}\right)+V_{2}\right]\epsilon/\hbar}\label{eq:PI_deriv-interm}
\end{equation}
The expression on the first line in the right hand side in the $n\rightarrow\infty$
limit (and hence when $\left|\epsilon\right|\rightarrow0$) becomes
the path integral associated with the action 
\[
S_{1}\left[x(z)\right]=\int_{z^{\prime}}^{z}L_{1}\left(z_{1}\right)dz_{1}=\int_{z^{\prime}}^{z}\left(K_{1}-V_{1}\left(z_{1}\right)\right)dz_{1}
\]
where the atomic velocities were defined with respect to the ``time''
$\epsilon\equiv\Delta z$ on the contour. Next, on the second line
in Eq. (\ref{eq:PI_deriv-interm}) we have an \emph{ordered }product
of exponential operators, with times on $\gamma$ increasing from
right to left from $z^{\prime}$ to $z$. In the $n\rightarrow\infty$
limit these could then be written as a time ordered exponent
\[
\widehat{U}_{2}\left(z,z^{\prime}\right)=\widehat{\mathcal{T}}_{\gamma}\exp\left\{ -\frac{i}{\hbar}\int_{z^{\prime}}^{z}\mathcal{H}_{2}\left(z_{1}\right)dz_{1}\right\} 
\]
where $\mathcal{H}_{2}\left(z\right)=K_{2}+V_{2}+V_{12}\left(z\right)$
and the integral is taken between $z^{\prime}$ and $z$ on the contour.
Therefore, we have just proved an exact identity 
\[
\left.\left\langle x_{f}\left|\widehat{U}_{\gamma}\left(z,z^{\prime}\right)\right|x_{0}\right.\right\rangle =\int_{x\left(z^{\prime}\right)=x_{0}}^{x\left(z\right)=x_{f}}\mathcal{D}x\left(z\right)e^{iS_{1}\left[x\left(z\right)\right]/\hbar}\widehat{U}_{2}\left(z,z^{\prime}\right)
\]

Clearly, this result is valid for the times $z$ and $z^{\prime}$
lying anywhere on $\gamma$. Subtleties associated with factorization
of the exponential operators in the $n\rightarrow\infty$ limit are
rigorously discussed e.g. in Ref. \cite{Schulman-book-PI}.

Consider now the three cases we are actually interested in. 

(i) Both times lie on the upper track: $z=t_{1}$ and $z^{\prime}=t_{0}^{+}$.
Then, $\epsilon=\left(t_{1}-t_{0}\right)/n>0$, and we recover Eqs.
(\ref{eq:ion-propagator-el-operator}) and (\ref{eq:electronic-propagator}).

(ii) Both times lie on the lower track: $z=t_{0}^{-}$ and $z^{\prime}=t_{1}$.
Then, $\epsilon=\left(t_{0}-t_{1}\right)/n<0$, and we obtain the
reverse time propagator:
\[
\left.\left\langle x_{0}\left|\widehat{U}\left(t_{0},t_{1}\right)\right|x_{1}\right.\right\rangle =\int_{x^{\prime}\left(t_{1}\right)=x_{1}}^{x^{\prime}\left(t_{0}\right)=x_{0}}\mathcal{D}x^{\prime}(t)e^{-iS_{1}\left[x^{\prime}(t)\right]/\hbar}\widehat{U}_{2}\left(t_{0},t_{1}\right)
\]
where 
\[
\widehat{U}_{2}\left(t_{0},t_{1}\right)=\widehat{\mathcal{T}}_{-}\exp\left\{ \frac{i}{\hbar}\int_{t_{0}}^{t_{1}}\left[\mathcal{H}_{2}\left(t^{\prime}\right)+\mathcal{H}_{12}\left(x^{\prime}\left(t^{\prime}\right)\right)\right]dt^{\prime}\right\} \equiv\widehat{U}_{2}^{\dagger}\left(t_{1},t_{0}\right)
\]
is the corresponding electronic propagator. 

(iii) Both times lie on the vertical track: $z=t_{0}^{-}-i\tau$ and
$z^{\prime}=t_{0}^{-}$. In this case $\epsilon=-i\tau/n=-i\Delta\tau$,
the atomic velocity $\Delta\overline{x}/\Delta z=i\Delta\overline{x}/\Delta\tau$
acquires an extra $i$, and hence the following identity results:
\[
\left.\left\langle x\left|\widehat{U}_{\gamma}\left(\tau,0\right)\right|x_{0}\right.\right\rangle =\int_{\overline{x}\left(0\right)=x_{0}}^{\overline{x}\left(\tau\right)=x}\mathcal{D}\overline{x}\left(\tau\right)e^{-S_{1}^{0}\left[\overline{x}\left(\tau\right)\right]/\hbar}\widehat{U}_{2}\left(\tau,0\right)
\]
where $S_{1}^{0}$ is the Euclidean action, and 
\[
\widehat{U}_{2}\left(\tau,0\right)=\overline{\mathcal{T}}\,\exp\left\{ -\frac{1}{\hbar}\int_{0}^{\tau}\left[\mathcal{H}_{2}^{0}+\mathcal{H}_{12}^{0}\left(\overline{x}\left(\tau^{\prime}\right)\right)\right]d\tau^{\prime}\right\} 
\]
is the corresponding imaginary time propagator. In particular, when
the initial Hamiltonian does not depend on the imaginary time and
$\tau\equiv\beta\hbar$, the representation of the exponential operator
$e^{-\beta\mathcal{H}^{0}}$ is recovered, Eqs. (\ref{eq:initial-DM-via-path-int})
and (\ref{eq:imag-time-propag-operator}).

The obtained expressions enable one, when read from left to right,
to replace the matrix elements of the propagation operators with appropriate
partial path integrals over nuclear trajectories. This was used in
section \ref{subsec:Influence-functional}. However, if read from
right to left, the same identities can be used to replace path integrals
with the corresponding matrix elements of time-ordered propagators.
In fact, this has been done in section \ref{subsec:Nuclei-only-reduced-DM}
when going from Eq. (\ref{eq:final-path-int-expr-for-DM-1}) to (\ref{eq:imaginary-time-propagator}),
albeit for a much simpler case when only region 1 was present and
hence there were no operators $\widehat{U}_{2}$ anymore.

\section*{Appendix B}

Here we shall consider the case when region $C$ that includes electron
orbitals interacting with nuclear displacements does not contain all
orbitals of the whole system. In particular, this case may correspond
to a molecule on a surface, when the electron-nuclear interaction
is included only for the molecule, or a case of a molecular junction,
where this interaction is only considered explicitly for the central
region between the leads. 

To illustrate the general idea, we shall consider a molecular junction
with the central region (region $C$) to which electrodes (leads)
are attached. We shall adopt the simplest one-particle approximation
for the leads, i.e. the following electronic Hamiltonian $\mathcal{H}_{2}$
for all times $t\geq t_{0}$ will be assumed:
\begin{equation}
\mathcal{H}_{2}(z)=\sum_{\alpha k}\epsilon_{\alpha k}(z)c_{\alpha k}^{\dagger}c_{\alpha k}+\sum_{nm}T_{nm}(z)c_{n}^{\dagger}c_{m}+\sum_{n}\sum_{\alpha k}\left[T_{\alpha k,n}(z)c_{n}^{\dagger}c_{\alpha k}+\text{h.c.}\right]+V_{C}\label{eq:electronic-Hamiltonian}
\end{equation}
where the first term describes the leads, the second - the central
system, and the third - interaction of the latter with the leads.
The last term corresponds to electronic correlation effects which
are only considered non-zero in the central region $C$, i.e. $V_{C}$
only contains operators from this region. Multiple leads are assumed
here designated by $\alpha$. The index $k$ numbers states of a particular
lead. 

The matrix elements in the Hamiltonian depend on $z$ due to the following
reasons. Firstly, the dependence on $z$ of $\epsilon_{\alpha k}(z)$
comes from the fact that a time-dependent bias may be applied in the
junction to the leads, i.e. each lead may be subjected to a particular
potential $\phi_{\alpha}(t)$ for $z$ lying on the horizontal tracks
of $\gamma$; on the vertical track, corresponding to the initial
preparation of the system, there is no bias applied. Secondly, the
$z$ dependence of the central region related matrix elements, $T_{nm}(z)$
and $T_{\alpha k,n}(z)=T_{n,\alpha k}^{*}(z)$, is entirely due to
the fact that nuclei positions depend on $z$. Namely, on the horizontal
tracks nuclei are clumped at $\left\langle x_{A}\right\rangle _{t}$
(and hence evolve in time), while their positions are set to the constant
values $x_{A}^{0}$ on the vertical track. 

The electronic Hamiltonian (\ref{eq:electronic-Hamiltonian}) can
conveniently be rewritten in a simplified form:
\begin{equation}
\mathcal{H}_{2}(z)=\sum_{ab}h_{ab}(z)c_{a}^{\dagger}c_{b}+V_{C}\label{eq:electronic-Hamiltonian-simple}
\end{equation}
where the summation is run over all orbitals of the entire system,
and 
\begin{equation}
h_{\alpha k,\alpha^{\prime}k^{\prime}}(z)=\delta_{\alpha\alpha^{\prime}}\left\{ \begin{array}{cc}
\delta_{kk^{\prime}}\left[\epsilon_{\alpha k}+\phi_{\alpha}(t)\right], & \text{if }z\in\text{horizontal track}\\
\delta_{kk^{\prime}}\left(\epsilon_{\alpha k}-\mu\right) & \text{if }z\in\text{vertical track}
\end{array}\right.\label{eq:electronic-h_ij-leads}
\end{equation}
\begin{equation}
h_{\alpha k,n}(z)=\left\{ \begin{array}{cc}
T_{\alpha k,n}(t), & \text{if }z\in\text{horizontal track}\\
T_{\alpha k,n}^{0}, & \text{if }z\in\text{vertical track}
\end{array}\right.\label{eq:electronic-h_ij-lead_C}
\end{equation}
\begin{equation}
h_{nm}(z)=\left\{ \begin{array}{cc}
T_{nm}(t), & \text{if }z\in\text{horizontal track}\\
T_{nm}^{0}-\mu\delta_{nm}, & \text{if }z\in\text{vertical track}
\end{array}\right.\label{eq:eletronic-H_ij-central}
\end{equation}
Here we included explicitly the electronic chemical potential for
the Hamiltonian on the vertical track in accordance with the initial
density matrix (\ref{eq:Rho-0}). As was mentioned, the time dependence
of the matrix elements $T_{nm}(t)$ and $T_{\alpha k,n}(t)=T_{n,\alpha k}^{*}(t)$
on the horizontal tracks comes from the nuclear positions which are
chosen as $\left\langle x_{A}\right\rangle _{t}$.

Correspondingly, the whole Hamiltonian governing the evolution of
the electronic GF is, therefore, 
\begin{equation}
H_{2}^{\lambda}(z)=\mathcal{H}_{2}(z)+\mathcal{H}_{12}^{\lambda}(z)=\sum_{ab}\left[h_{ab}(z)+\lambda V_{ab}(z)\right]c_{a}^{\dagger}c_{b}+V_{C}\label{eq:tot-electronic-Hamiltonian-simple}
\end{equation}
Hence, the GF introduced above satisfies the usual equations of motion
based on this Hamiltonian:
\begin{equation}
i\hbar\partial_{z}G_{ab}^{\lambda}\left(z,z^{\prime}\right)=\delta_{ab}\delta\left(z-z^{\prime}\right)+\sum_{c}\left[h_{ac}(z)+\lambda V_{ac}(z)\right]G_{cb}^{\lambda}\left(z,z^{\prime}\right)+\int_{\gamma}dz_{1}\,\sum_{d\in C}\widetilde{\Sigma}_{ad}\left(z,z_{1}\right)G_{db}^{\lambda}\left(z_{1},z^{\prime}\right)\label{eq:EoM-for-GF_1}
\end{equation}
\begin{equation}
-i\hbar\partial_{z^{\prime}}G_{ab}^{\lambda}\left(z,z^{\prime}\right)=\delta_{ab}\delta\left(z-z^{\prime}\right)+\sum_{c}G_{ac}^{\lambda}\left(z,z^{\prime}\right)\left[h_{cb}(z^{\prime})+\lambda V_{cb}\left(z^{\prime}\right)\right]+\int_{\gamma}dz_{1}\,\sum_{d\in C}G_{ad}^{\lambda}\left(z,z_{1}\right)\widetilde{\Sigma}_{db}\left(z_{1},z^{\prime}\right)\label{eq:EoM-for-GF_2}
\end{equation}
where 
\begin{equation}
\delta\left(z-z^{\prime}\right)=\frac{d}{dz}\theta_{zz^{\prime}}=\left\{ \begin{array}{cc}
\delta\left(t-t^{\prime}\right), & \text{if }z,z^{\prime}\in\text{upper track}\\
-\delta\left(t-t^{\prime}\right), & \text{if }z,z^{\prime}\in\text{lower track}\\
i\delta\left(\tau-\tau^{\prime}\right) & \text{if }z,z^{\prime}\in\text{vertical track}\\
0, & \text{in all other cases}
\end{array}\right.\label{eq:delta-function-Def}
\end{equation}
which satisfies the usual filtering theorem on the contour $\gamma$
for any ``good'' function $f(z)$:
\[
\int_{\gamma}f\left(z^{\prime}\right)\delta\left(z-z^{\prime}\right)dz^{\prime}=f(z)
\]
and we have introduced an electronic self-energy matrix $\widetilde{\Sigma}_{CC}\left(z,z^{\prime}\right)$
defined only for orbitals in the central region. Note that the last
term in Eq. (\ref{eq:EoM-for-GF_1}) is only kept for $a\in C$, while
in Eq. (\ref{eq:EoM-for-GF_2}) it survives for $b\in C$. 

Writing Eq. (\ref{eq:EoM-for-GF_1}) in the matrix form for the blocks
$CC$ and $\alpha C$, solving for $\mathbf{G}_{\alpha C}^{\lambda}$
and substituting it into the equation for $\mathbf{G}_{CC}^{\lambda}$,
one obtains using the matrix notations:
\begin{equation}
i\hbar\partial_{z}\mathbf{G}_{CC}^{\lambda}\left(z,z^{\prime}\right)=\mathbf{1}_{C}\delta\left(z-z^{\prime}\right)+\left[\mathbf{h}_{C}(z)+\lambda\mathbf{V}_{C}(z)\right]\mathbf{G}_{CC}^{\lambda}\left(z,z^{\prime}\right)+\int_{\gamma}dz_{1}\,\overline{\Sigma}_{CC}\left(z,z_{1}\right)\mathbf{G}_{CC}^{\lambda}\left(z_{1},z^{\prime}\right)\label{eq:G_CC_with_corr}
\end{equation}
where $\overline{\Sigma}_{CC}\left(z,z_{1}\right)=\Sigma_{CC}\left(z,z_{1}\right)+\widetilde{\Sigma}_{CC}\left(z,z_{1}\right)$
is the composed self-energy, containing, apart from the correlation
component, also the self-energy of the electrodes \cite{Stefanucci-Leeuwen}:
\begin{equation}
\varSigma_{CC}\left(z,z^{\prime}\right)=\sum_{\alpha}\varSigma_{CC}^{\alpha}\left(z,z^{\prime}\right)=\sum_{\alpha}\mathbf{h}_{C\alpha}\left(z\right)\mathbf{g}_{\alpha}\left(z,z^{\prime}\right)\mathbf{h}_{\alpha C}\left(z^{\prime}\right)\label{eq:T1-selfenergy}
\end{equation}
Here $\mathbf{g}_{\alpha}\left(z,z^{\prime}\right)$ is the GF of
the isolated lead $\alpha$ associated with the Hamiltonian $\mathbf{h}_{\alpha}(z)$.
We have assumed that the leads do not directly interact. It is seen
that the leads and correlation self-energies simply add up. 

In the same way one can define the ``unperturbed'' GF, $\mathbb{G}\left(z,z^{\prime}\right)$,
defined without the electron-nuclear coupling (section \ref{subsec:Green's-function});
it satisfies
\begin{equation}
i\hbar\partial_{z}\mathbb{G}_{CC}\left(z,z^{\prime}\right)=\mathbf{1}_{C}\delta\left(z-z^{\prime}\right)+\mathbf{h}_{C}(z)\mathbb{G}_{CC}\left(z,z^{\prime}\right)+\int_{\gamma}dz_{1}\,\overline{\Sigma}_{CC}\left(z,z_{1}\right)\mathbb{G}_{CC}\left(z_{1},z^{\prime}\right)\label{eq:EoM-for-zer-GF}
\end{equation}
It is essential that $\mathbb{G}\left(z,z^{\prime}\right)$ is defined
by the matrix $\mathbf{h}(z)$, which is the same on both horizontal
tracks of $\gamma$. Note that, because of the time dependence of
the nuclear positions on the horizontal tracks given by $\left\langle x_{A}\right\rangle _{t}$,
the calculation of $\mathbb{G}\left(z,z^{\prime}\right)$ might be
non-trivial. If such a dependence is ignored (and, e.g. replaced by
the equilibrium positions $x_{A}^{0}$), then this GF can be calculated
explicitly e.g. in the wide band approximation and in absence of electronic
correlation \cite{Ridley-current-PRB-2015,Ridley-current-PRB-2016,Ridley-current-PRB-2017}. 

The two functions $\mathbf{G}_{CC}^{\lambda}$ and $\mathbb{G}_{CC}$
are related by a Dyson-type equation. To derive it, we rewrite Eqs.
(\ref{eq:G_CC_with_corr}) and (\ref{eq:EoM-for-zer-GF}) in a symbolic
form to keep simple notations:
\begin{equation}
\left(i\hbar\partial_{z}-\mathbf{h}_{z}\right)\mathbf{G}^{\lambda}=\mathbf{1}\delta+\lambda\mathbf{V}\mathbf{G}^{\lambda}+\overline{\Sigma}\mathbf{G}^{\lambda}\label{eq:G-symbolic}
\end{equation}
\begin{equation}
\left(i\hbar\partial_{z}-\mathbf{h}_{z}\right)\mathbb{G}=\mathbf{1}\delta+\overline{\Sigma}\mathbb{G}\label{eq:G0-symbolic}
\end{equation}
Here convolutions are assumed in a product of any two-time quantities,
e.g. in $\overline{\Sigma}\mathbf{G}^{\lambda}$. If we introduce
the GF for the central region, $\mathbf{g}_{C}\left(z,z^{\prime}\right)$
(to be denoted simply by $\mathbf{g}$ in our symbolic notations),
which satisfies $\left(i\hbar\partial_{z}-\mathbf{h}_{z}\right)\mathbf{g}\mathbf{=\mathbf{1}\delta}$,
then the two equations transform into: 
\[
\mathbf{G}^{\lambda}=\mathbf{g}+\lambda\mathbf{g}\mathbf{V}\mathbf{G}^{\lambda}+\mathbf{g}\overline{\Sigma}\mathbf{G}^{\lambda}
\]
\[
\mathbb{G}=\mathbf{g}+\mathbf{g}\overline{\Sigma}\mathbb{G}
\]
From the second equation $\mathbf{g}=\left(\mathbf{1}-\mathbf{g}\overline{\Sigma}\right)\mathbb{G}$,
which, when used in the first, gives the required relationship, 
\[
\mathbf{G}^{\lambda}=\mathbb{G}+\lambda\mathbb{G}\mathbf{V}\mathbf{G}^{\lambda}
\]
which has exactly the form of the Dyson equation (\ref{eq:T1-Dayson-eq}). 

Hence, in any case when the region $C$ does not cover the whole system,
the contribution of the rest of the system that does not interact
directly with nuclear displacements in region 1 manifests itself in
the properly defined self-energy which is simply added to the correlation
self-energy of region $C$. Therefore, having this in mind, one can
use only the block $CC$ of the Green's Functions in the actual calculations.

\section*{Appendix C}

Calculation of the contour integral in Eq. (\ref{eq:GF-to-the-1st-order})
requires a generalization of the Langreth rules \cite{Stefanucci-Leeuwen}
for the case when the Hamiltonian on different horizontal tracks is
different, and there is an extra single variable function in the convolution,
$\mathbf{V}\left(z\right)$. Only the lesser component is needed for
our purposes here; however, we have to consider the cases of $z$
being on the upper, lower and vertical tracks.

Therefore, let us consider an integral over the contour:
\begin{equation}
C\left(z,z^{\prime}\right)=\int_{\gamma}dz_{1}\,A\left(z,z_{1}\right)V\left(z_{1}\right)B\left(z_{1},z^{\prime}\right)\label{eq:Def-convoluiton}
\end{equation}
Performing the integration explicitly over each of the three tracks
on $\gamma,$ the following identities can be established:
\[
C_{++}^{<}\left(t,t^{\prime}\right)=-i\int_{0}^{\beta\hbar}d\tau\,A_{+M}V_{M}B_{M+}
\]
\begin{equation}
+\int_{t_{0}}^{t_{1}}dt\left(A_{++}^{r}V_{+}B_{++}^{<}+A_{++}^{<}V_{+}B_{++}^{a}+A_{++}^{<}V_{+}B_{++}^{>}-A_{+-}V_{-}B_{-+}\right)\label{eq:C-lesser_++}
\end{equation}
\[
C_{--}^{<}\left(t,t^{\prime}\right)=-i\int_{0}^{\beta\hbar}d\tau\,A_{-M}V_{M}B_{M-}
\]
\begin{equation}
+\int_{t_{0}}^{t_{1}}dt\left(A_{--}^{r}V_{-}B_{--}^{<}+A_{--}^{<}V_{-}B_{--}^{a}+A_{-+}V_{+}B_{+-}-A_{--}^{>}V_{-}B_{--}^{<}\right)\label{eq:C-lesser_--}
\end{equation}
\[
C_{MM}^{<}\left(\tau,\tau^{\prime}\right)=\int_{t_{0}}^{t_{1}}dt\left(A_{M+}V_{+}B_{+M}-A_{M-}V_{-}B_{-M}\right)
\]
\begin{equation}
-i\int_{0}^{\beta\hbar}d\tau_{1}\left(A_{MM}^{r}V_{M}B_{MM}^{<}+A_{MM}^{<}V_{M}B_{MM}^{a}+A_{MM}^{<}V_{M}B_{MM}^{>}\right)\label{eq:C-lesser_MM}
\end{equation}
Above, for simplicity of notations, the arguments of the functions
are omitted; that should not cause any confusion as all integrals
correspond to convolutions. Also, note that the retarded and advanced
components can only be introduced when both times are positioned on
the same track (either forward, backward or vertical). 

Usual Langreth rules \cite{Stefanucci-Leeuwen} are recognized here
when there is no difference between the two horizontal tracks and
the function $V(z)\equiv1$. For instance, consider the first formula
(\ref{eq:C-lesser_++}): in this case there is no difference between
$A_{+-}=A^{<}$ and $A_{++}^{<}$, $V_{+}=V_{-}\equiv1$ and also,
$B_{++}^{>}=B_{-+}$, so that the third and the fourth contributions
in the real time integral cancel out, and we arrive at the usual rules
for the lesser function. Also, both expressions (\ref{eq:C-lesser_++})
and (\ref{eq:C-lesser_--}) give identical results.


\end{document}